\documentclass[twocolumn,showpacs,preprintnumbers,amsmath,amssymb]{revtex4} 

\usepackage{graphicx}% Include figure files                                                                                                 
\usepackage{dcolumn}% Align table columns on decimal point                                                                                 
\usepackage{bm}% bold math    
\usepackage{psfrag}
\usepackage{subfigure}

\begin{document}

%\preprint{APS/123-QED}

\title{Exploring Parameter Constraints on Quintessential Dark Energy: The Exponential Model\\}    

\author{Brandon Bozek, Augusta Abrahamse,  Andreas Albrecht, and Michael Barnard}                             
\affiliation{Physics Department, University of California, Davis.}                                                                                                                                                                                                                                               
\date{\today}                                                                                               

\begin{abstract}
We present an analysis of a scalar field model of dark energy  with an exponential
potential using the Dark Energy Task Force (DETF) simulated data
models. Using Markov Chain Monte Carlo sampling techniques we examine
the ability of each simulated data set to constrain the parameter
space of the exponential potential for data sets based on a cosmological constant 
and a specific exponential scalar field model.  We compare our results
with the constraining power calculated by the DETF using their
``$w_0-w_a$'' parametrization of the dark energy.  We find that
respective increases in constraining power from one stage to the next 
produced by our analysis give results consistent with DETF results. To further
investigate the potential impact of future experiments, we also
generate simulated data for an exponential model background cosmology
which can not be distinguished from a cosmological constant at DETF 
``Stage 2'', and show that for this cosmology good DETF Stage 4 data 
would exclude a cosmological constant by better than 3$\sigma$.
\end{abstract}

\maketitle

\section{\label{sec:level1}Introduction}

In the late 90's, two independent teams presented evidence from
supernova observations that the universe, instead of slowing down due
to gravity, is accelerating \cite{Riess:1998cb, Perlmutter:1998np}. In the standard cosmological framework, the
acceleration is caused by a mysterious new form of matter, dubbed  
``Dark Energy'', that makes up roughly 70\% of the
universe. There is a wide variety of possible explanations for dark
energy. The simplest model that provides a good fit to the data is a
cosmological constant. A cosmological constant is equivalent to a
homogeneous fluid with a constant energy density and a ratio of pressure to energy density  (the
``equation of state parameter'' $w$), equal to $-1$ at all times. Yet, despite
compelling evidence for the existence of dark energy, it is unclear
whether the dark energy density is constant or varies with 
time. There are many different proposals for a dynamical form of dark
energy, one of them being quintessence. Quintessence describes the
acceleration being caused by a scalar field, $\phi$, but even just
among quintessence models there is a tremendous variety of possible
behaviors.  There is considerable interest in acquiring better data in
order to improve our understanding of dark energy.

Recently the Dark Energy Task Force (DETF) released a report
charting a course for future experiments \cite{Albrecht:2006um}.  They modeled dark energy as a
homogeneous and isotropic fluid with an equation of state
parameterized by $w(a) = w_0 + w_a(1-a)$, where the scale factor $a =
1$ today. Defining Stage 1 to be what is already known, they
forecasted data for three additional experimental stages: Stage 2 data
represents on-going experiments that will be completed 
in the near future. Stage 3 data sets represent medium sized proposed
experiments. Lastly, Stage 4 data sets represent proposed
large scale future space and ground-based experiments. Each stage is
further categorized as either ``optimistic'' or  
``pessimistic'' depending on how well the systematics are expected to be
constrained. The scientific impact of a stage was quantified in terms
of a  ``Figure of Merit'' (FoM). The Figure of Merit is defined to be
the ratio of the area of the 2$\sigma$ contour of the $w_0 - w_a$
space for Stage 2 divided by the area of the 2$\sigma$ contour of the
$w_0 - w_a$ space for Stage 3 (or 4).

\begin{figure*}[!t]
\begin{center}
\includegraphics[trim = 0mm 50mm 0mm 40mm, clip,width=0.45\textwidth]{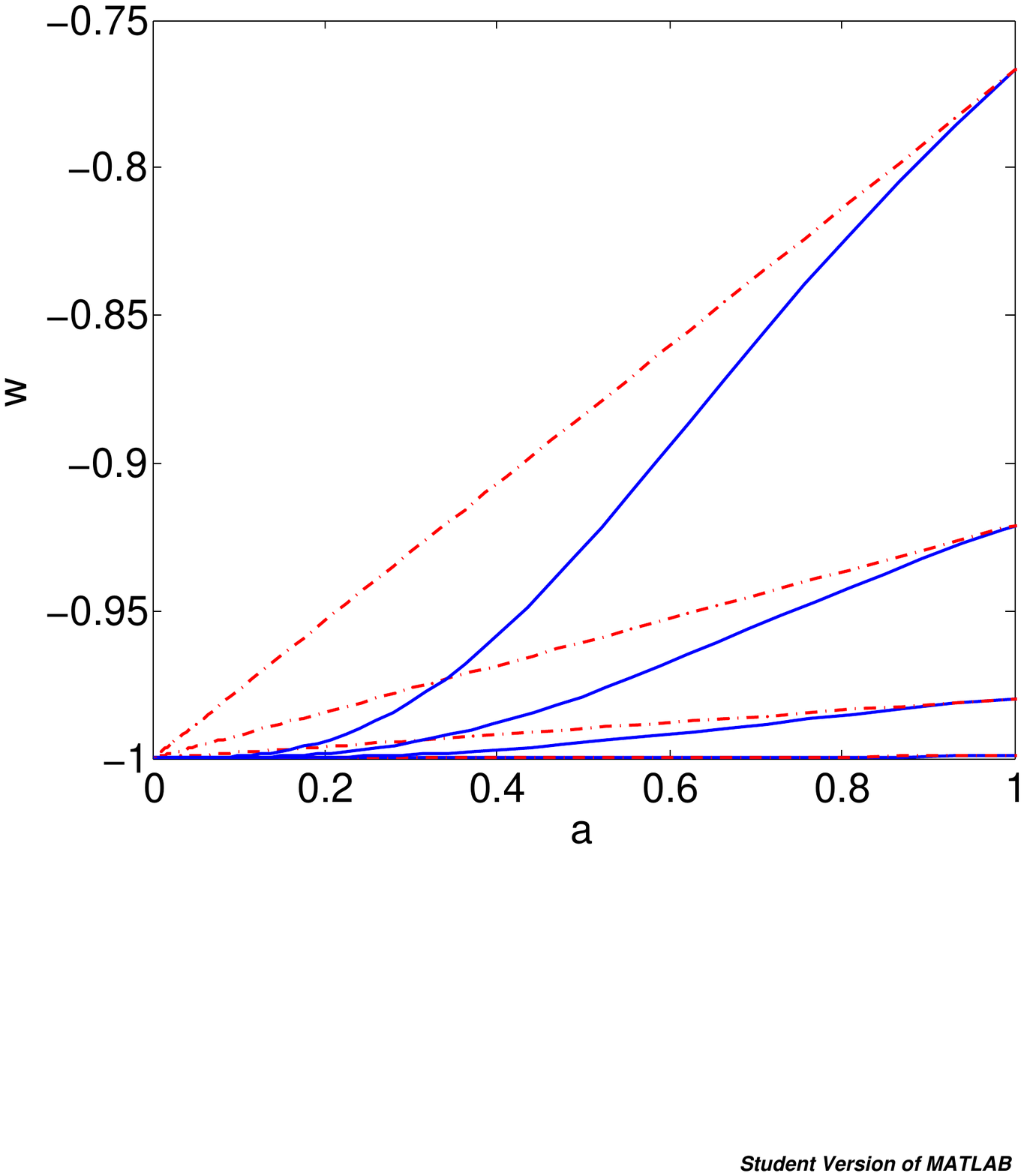}
\includegraphics[trim = 0mm 50mm 0mm 40mm, clip, width=0.45\textwidth]{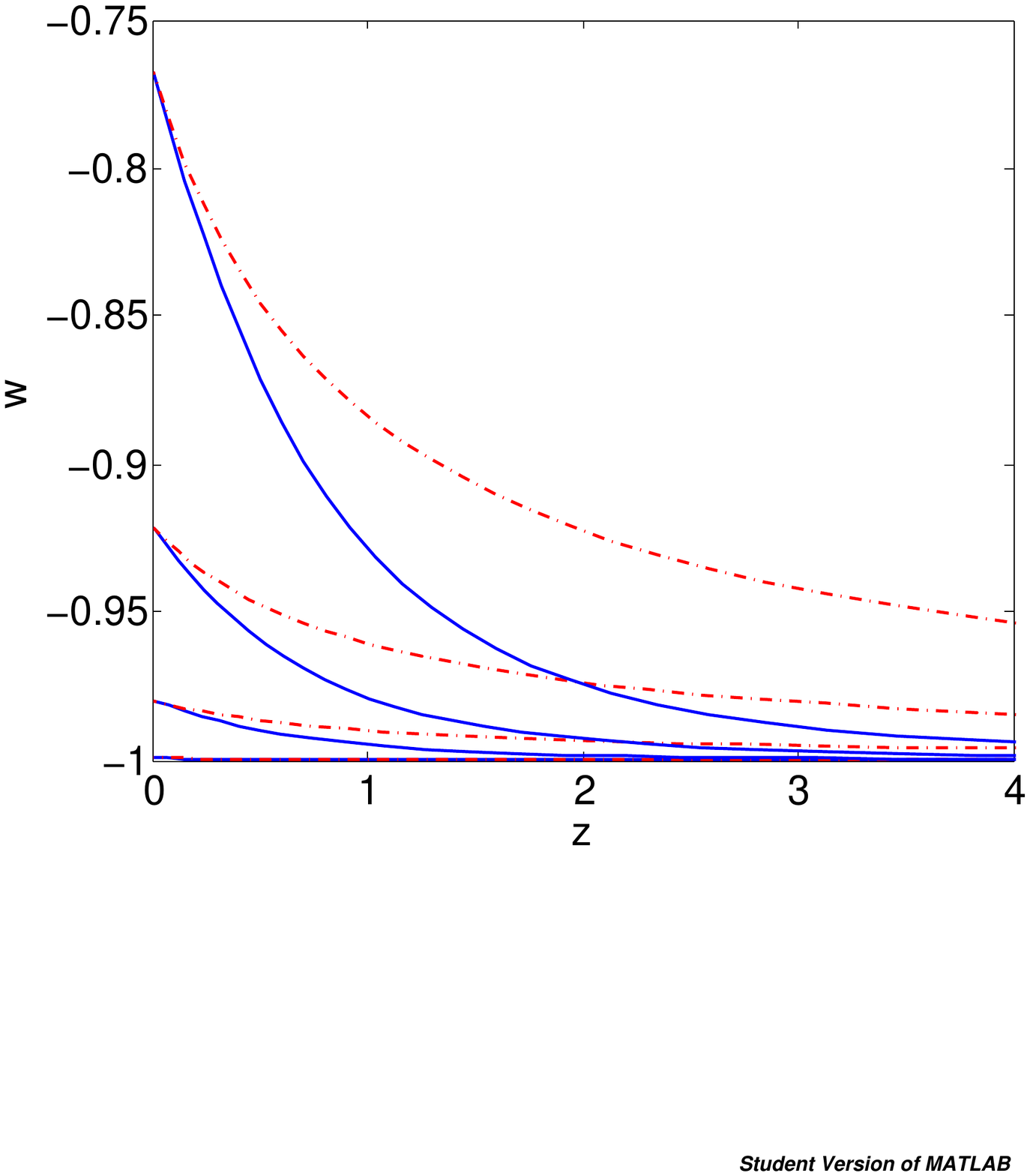}

\caption{Here we illustrate the differences between $w(a)$ curves from the
  exponential model (solid) and from the ansatz used by the DETF (dot-dashed). Curves
are plotted with respect to $a$ (left panel) and 
  $z$ (right panel). The four curves from top to bottom are given by $(w_0,w_a)$:
  $(-0.77673 ,-0.2327)$, $(-0.9211, -0.0789)$, $(-0.9797, -0.0203)$,
  $(-0.9992, -0.0008)$.  Using the final parameterization (explained in Sect. \ref{sec:Sec3}), the 
exponential model parameters are given by $(\lambda, V_I)$: $(0.07, 0.3725)$,
  $(0.35, 0.42)$, $(0.7, 0.42)$, $(1.2, 0.52)$ (in units defined in
  the text). Both models give
  $w(a=1)$ to be the same value for each set of curves and $w(a_I) =
  -1$ for all curves shown.} 
\label{fig:w0wacomp}
\end{center}
\end{figure*}

The DETF  analysis leaves several open questions, some of which our research seeks to
address. The $w_0-w_a$ parameterization is not motivated by a physical
model of dark energy and provides cosmological solutions that may be
very different from a scalar field model. As illustrated in Fig.
\ref{fig:w0wacomp}, the $w$ curves generated by
the exponential scalar field model that we consider in this paper are
not especially well fit by curves in the $w_0-w_a$ family,
except for those nearly identical to $w = -1$.  Thus the relationship
between the DETF results and the impact of future experiments on
scalar field models is not clear.  A good way to clarify this point is to
model the impact of future data sets directly on particular scalar
field quintessence models, which is what we do here. Our work 
complements the DETF report as well as work by other authors 
using alternative $w(a)$ parameterizations \cite{Liddle:2006kn,Albrecht:2007qy},
parameterizations of $\rho_{DE}$ \cite{Dick:2006ev}, and model
independent scalar field parameters \cite{Huterer:2006mv}.  

The exponential scalar field model has been used in many different cosmological
contexts due to its ability to give scaling solutions for the scalar
field energy density $\rho_\phi$ where
$\frac{d(log(\rho_{\phi}))}{d(log(a))} \rightarrow \beta$. The
constant $\beta$ depends on parameters in the scalar field potential
as well as the other forms of matter present in the universe . Originally the
potential was used 
for power law inflation models and was shown to have a range of
attractor solutions \cite{Halliwell:1986ja}. Its ability to produce
attractor solutions that scale like the background energy density made
it an interesting choice for a dark matter candidate
\cite{Ferreira:1997hj, Ferreira:1997au, Ratra:1987rm}.  The
variety of scaling solutions is well covered by Copeland, et
al. \cite{Copeland:1997et} where they argue that consideration of ``fine
tuning''parameters and constraints from nucleosynthesis give $\lambda
> 20$ as a natural choice. This range of $\lambda$ values wouldn't allow for late time
cosmological acceleration and is therefore ruled out as an explanation
for dark energy. The ``fine tuning'' that is required to
successfully describe dark energy with this model is needed so that 
the scalar field energy density can be initially very small; of
the order of the dark energy density today. This has caused this model to
be discarded by many authors on the basis that the model has lost the
theoretical generality that made the potential initially
interesting. However, as a practical matter the fine tuning is
straightforward to implement, and the potential is very simple and
easy to work with.  Because of this simplicity, we found it valuable
to have this potential as part of our larger project (which includes a
variety of more complicated quintessence potentials \cite{Abrahamse:2007ah, Barnard:2007ah}). The
simplicity helped us deal with a number of technical issues first with
the exponential model and then transfer our understanding to the more
complicated cases. 
In addition, realistic cosmologies for the exponential model have
their special forms for $w(a)$ (illustrated in
Fig. \ref{fig:w0wacomp}). We found it useful to include this family of $w(a)$ curves in our set of
possibilities to more fully understand the constraining power of future
data sets.

The paper is organized as follows. In Section \ref{sec:Sec2} we provide an
introduction to our scalar field model and its cosmological
solutions. In Section \ref{sec:Sec3} we describe how we come about our choice of
parameterization, as this is a critical step in the MCMC
analysis. (An account of our general MCMC methods and data modeling can be found in the
appendix of our companion paper \cite{Abrahamse:2007ah}.  This paper contains
only information specific to the exponential model.)  Section
\ref{sec:Sec4} presents our results for data simulated using a
background cosmology with a cosmological constant and then Section
\ref{sec:Sec5} presents results where the data is based on a cosmology
with exponential model quintessence. 
Finally, we summarize our 
key results in the conclusions. 

\begin{figure*}[!t]
\begin{center}
\includegraphics[trim = 0mm 50mm 0mm 40mm, clip,width=0.45\textwidth]{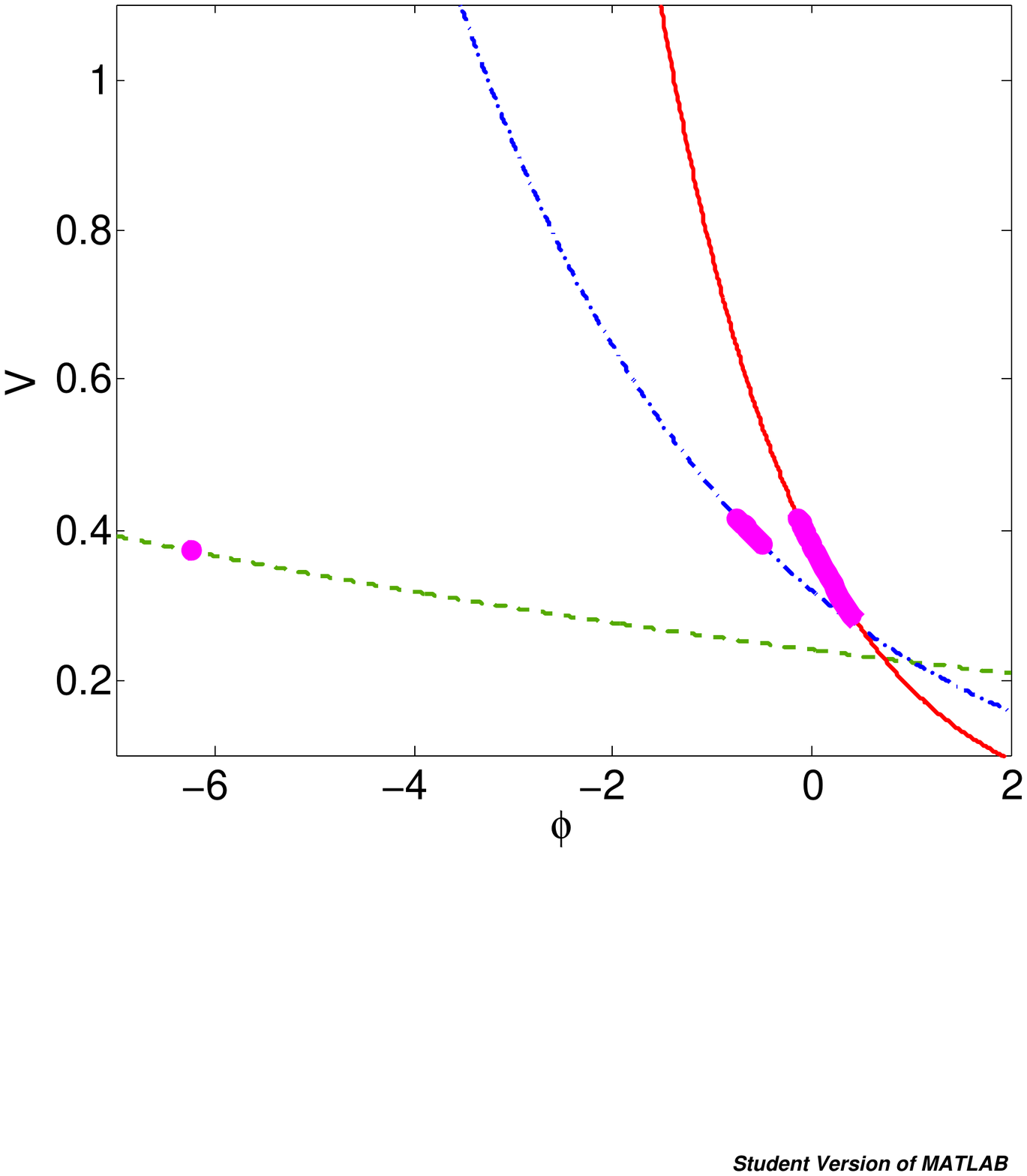} \quad
\includegraphics[trim = 0mm 50mm 0mm 40mm, clip, width=0.45\textwidth]{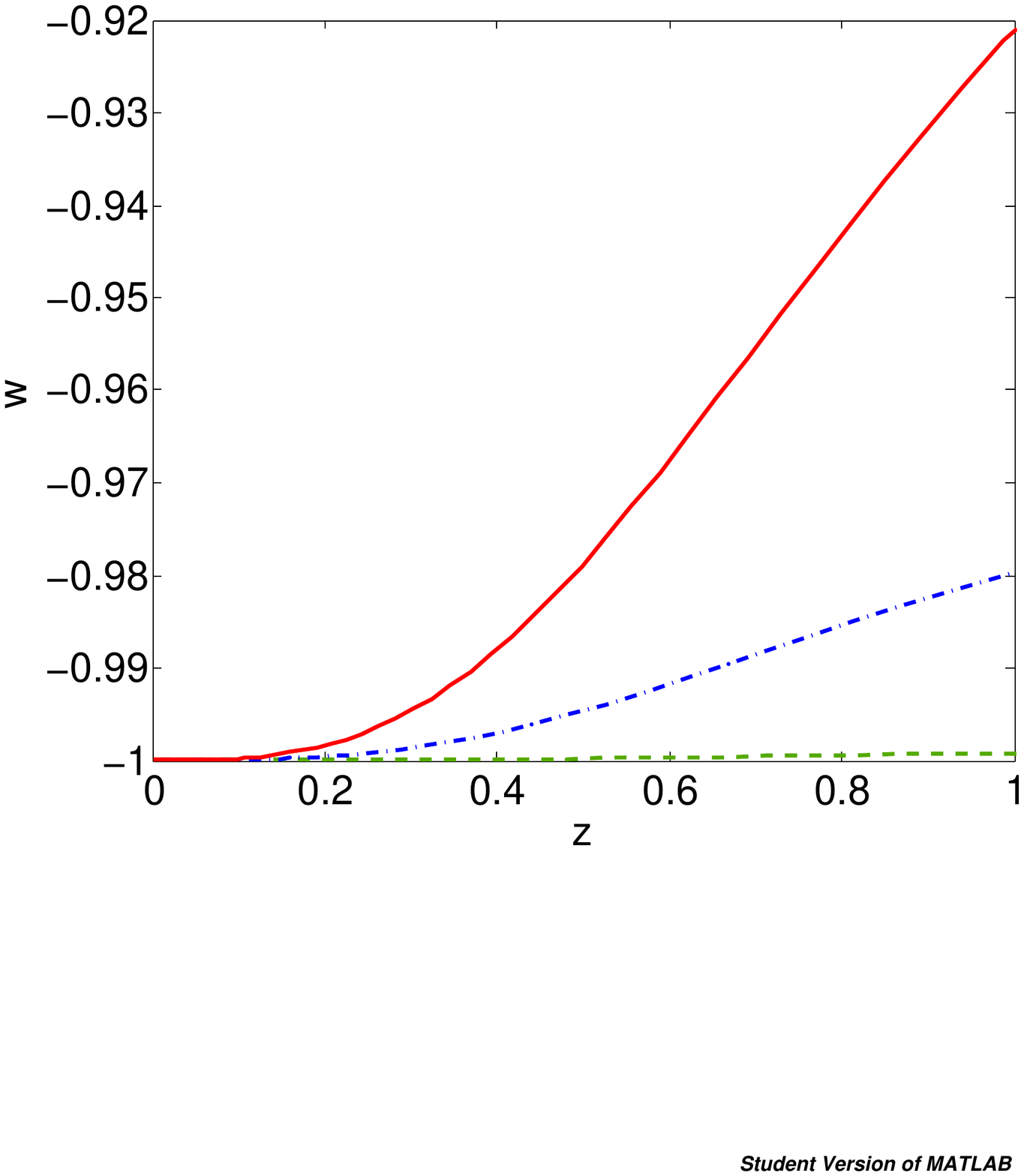}

\caption{The left panel shows three examples of the exponential
  potential (dashed line: $\lambda = 0.07, V_0 = 0.24, \phi_I =
  -6.28$, long-short dashed line: $\lambda = 0.35, V_0 = 0.32,
  \phi_I = -0.78$, solid line: $\lambda = 0.7, V_0 = 0.38, \phi_I
  = -0.14$. $V_0$ is given in units of $h^2$, as mentioned in Section
  \ref{sec:Sec4}). The path of the scalar field is depicted by thick
  solid curves. The corresponding $w$ behavior is shown in the right
  panel. The solid curve gives the potential used for the fiducial
  model discussed in Section \ref{sec:Sec5}.} 
\label{fig:firstpot}
\end{center}
\end{figure*}

\section{\label{sec:Sec2}Exponential Model Cosmology}
We model dark energy as a homogeneous scalar field evolving in an exponential potential
\begin{equation}
V = V_0 e^{-\lambda\phi}.
\end{equation}
The cosmological evolution of this scalar field in a FRW universe is then given by solving:
\begin{equation}
\frac{d^2\phi}{dt^2} + 3H\frac{d\phi}{dt} + \frac{dV(\phi)}{d\phi} = 0
\end{equation}
\begin{equation}
H^2 = \frac{1}{3 M_{p}^2}(\rho_r + \rho_m + \rho_\phi) - \frac{k}{a^2}
\end{equation}
\begin{equation}
\rho_\phi = \frac{1}{2}(\frac{d\phi}{dt})^2 + V(\phi)
\end{equation}
where $M_{p}$ is the reduced Planck mass.
The equation of state of the scalar field is given by 
\begin{equation}
w = \frac{\frac{1}{2}(\frac{d\phi}{dt})^2 - V(\phi)}{\frac{1}{2}(\frac{d\phi}{dt})^2 + V(\phi)},
\end{equation}
which we will use in discussing an evolving dark energy.  In this
picture a cosmological constant is equivalent to a scalar field with $w = -1$.

In our analysis we initially set $\frac{d\phi}{dt} = 0\ $\footnote{We
  set the initial value of  $\frac{d\phi}{dt}$ to zero because it
  simplifies the problem while still giving us an interesting set of
  solutions to work with. Allowing nonzero initial values of
  $\frac{d\phi}{dt}$ would increase the fine tuning issues without
  leading to a more interesting set of cosmologies.}. This leaves the
  dynamics of the field completely determined by the slope and
  curvature of the potential: 
\begin{equation}
\frac{dV}{d\phi} = -\lambda V_0 e^{-\lambda\phi} , \frac{d^2V}{d\phi^2} = \lambda^2 V_0 e^{-\lambda\phi}
\end{equation}
Since the initial field velocity is zero, the initial equation of
state is $w = -1$, mimicking a cosmological constant. As the field begins
to roll the equation of state begins to depart from $-1$. The rate of
this departure is determined by the steepness of the potential. 
A steeper slope
gives changes in $\phi$ that correspond to larger changes in
$w$. Likewise, a flat slope gives little change in $\phi$ and therefore has a
cosmology similar to a cosmological constant, as shown in Figure
\ref{fig:firstpot}.

The  possible scaling solutions achievable by the exponential model
  are systematically discussed by Copeland, et
  al. \cite{Copeland:1997et}. If a scaling solution is reached before
  the onset of dark energy domination the universe will not accelerate
  and therefore is a poor match to current data. However, there are
  subsets of scaling solutions that reach their scaling solution after
  dark energy domination and provide different fates for the
  universe. These scaling solutions fall into two categories: (i) $ 0 <
  \lambda < \sqrt{2}$ or (ii) $\sqrt{2} \le \lambda 
< \lambda_{\ast}$. Solutions with $\lambda$ values in category (i) approach
scaling where $w \rightarrow \frac{\lambda^2}{3} - 1$, giving
late time acceleration. For category (ii), we define
$\lambda_{\ast}$ to be the value that gives $w(a = 1) <
-\frac{1}{3}$, but the scaling solution leads to $w(a >1) >
-\frac{1}{3}$. It is possible for $\lambda_{\ast}$ to be larger
than $\sqrt{3}$ and therefore have $w(a >1) \rightarrow 0$. This value
depends on the initial scalar field energy density, $\rho_{\phi,I}$, which in turn determines when the
field begins to approach its scaling solution, i.e. when the field
starts rolling. We are allowing $\rho_{\phi,I}$ and other cosmological
parameters to float so a universal value of $\lambda_{\ast}$ cannot be
uniquely determined. 
 
\section{\label{sec:Sec3}Parametrization}

\begin{figure*}[!t]
\centering
%\vspace{-80pt}
\includegraphics[trim = 0mm 50mm 0mm 40mm, clip,width=0.45\textwidth]{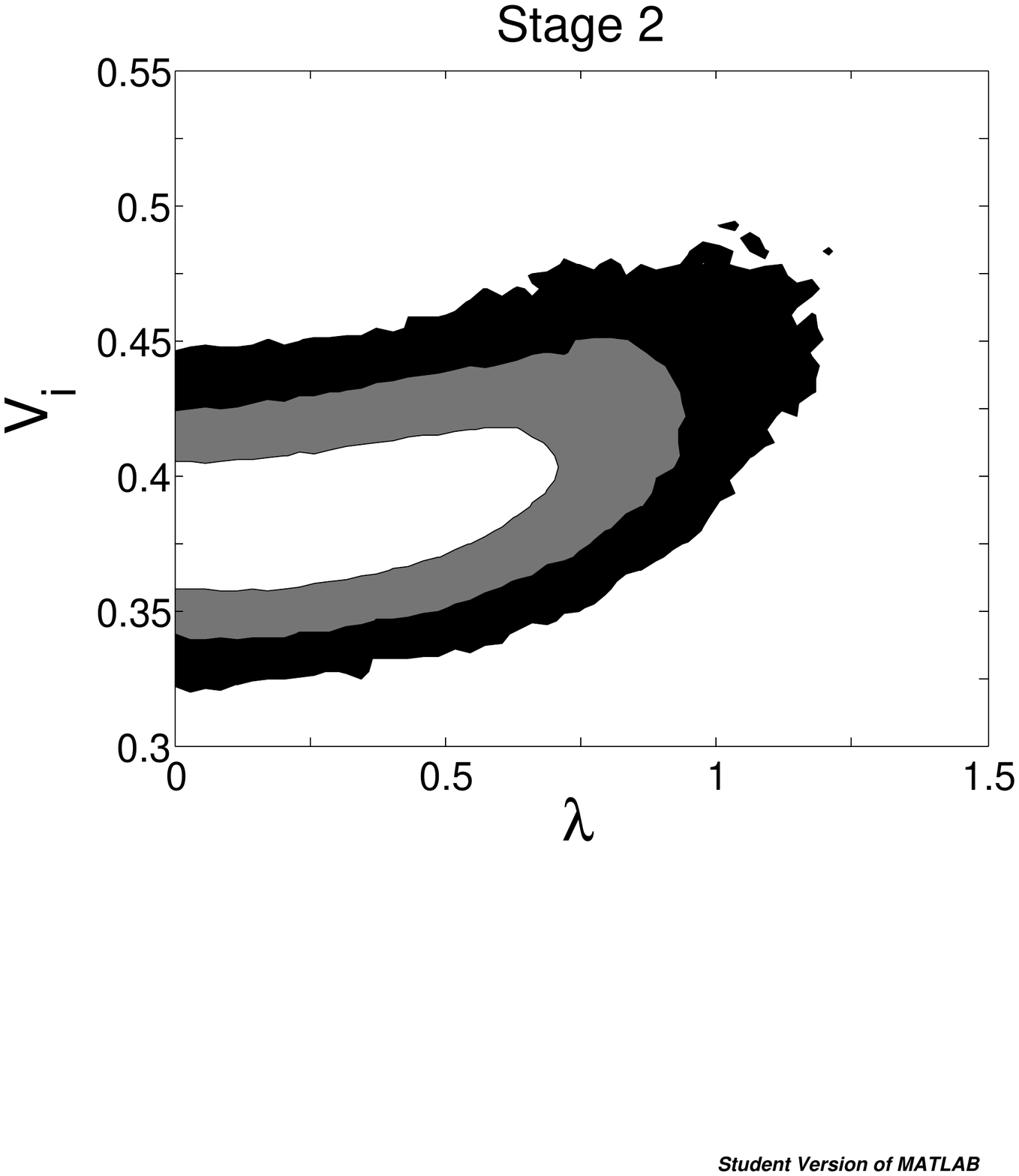}
\vspace{-20pt}
\includegraphics[trim = 0mm 50mm 0mm 40mm, clip,width=0.45\textwidth]{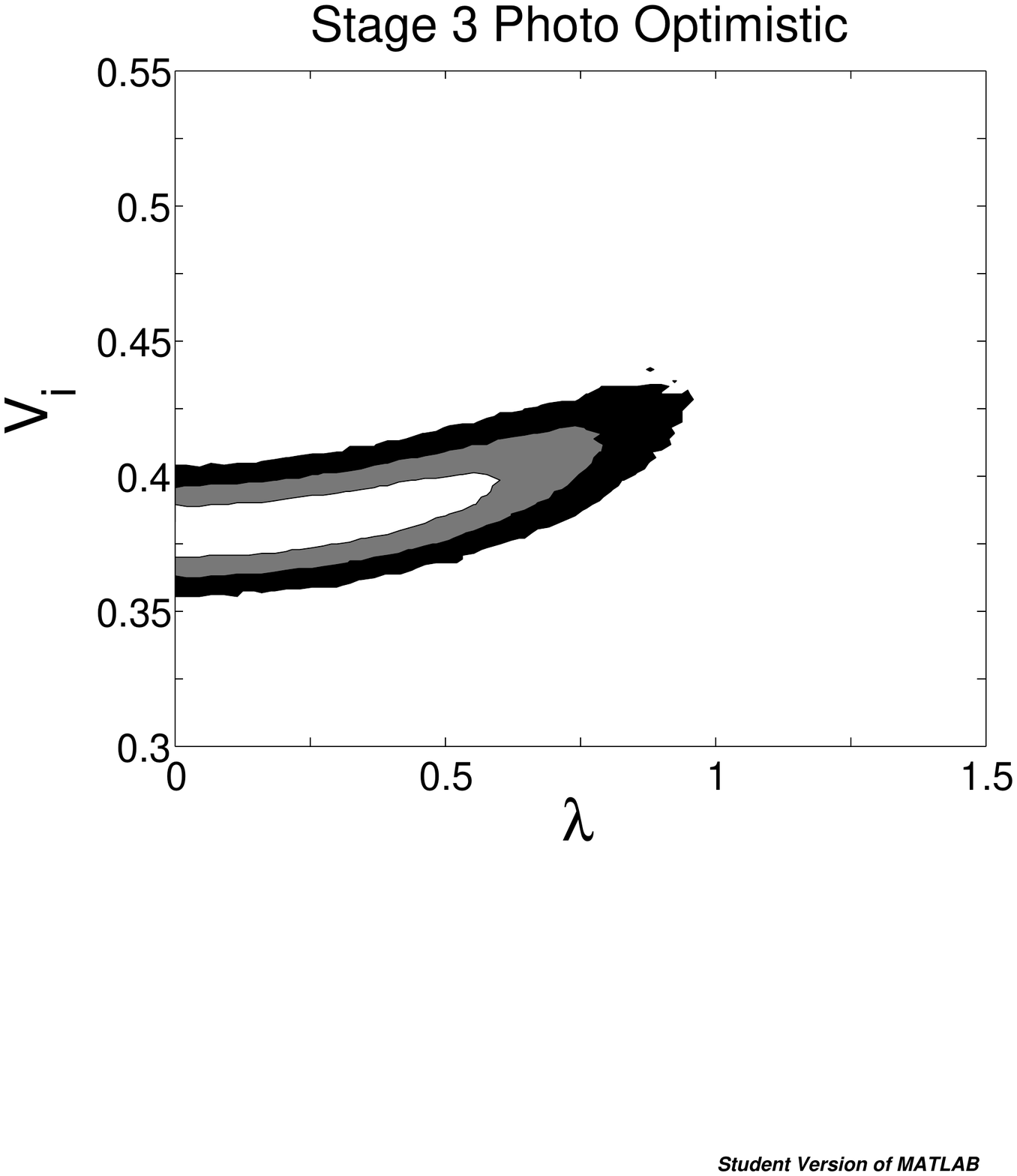}
\vspace{-20pt}
\includegraphics[trim = 0mm 50mm 0mm 40mm, clip,width=0.45\textwidth]{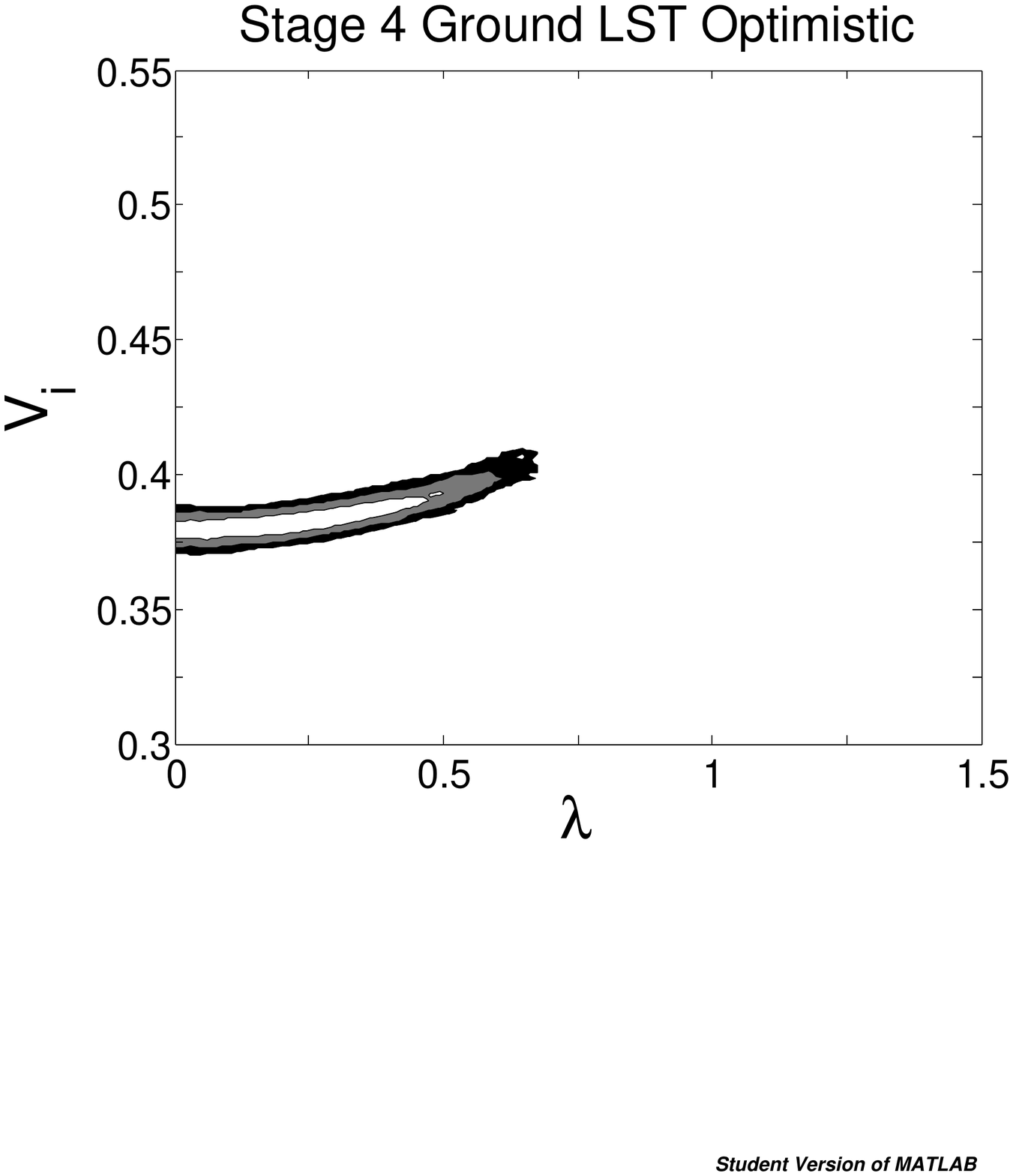}
\vspace{-20pt}
\includegraphics[trim = 0mm 50mm 0mm 40mm, clip,width=0.45\textwidth]{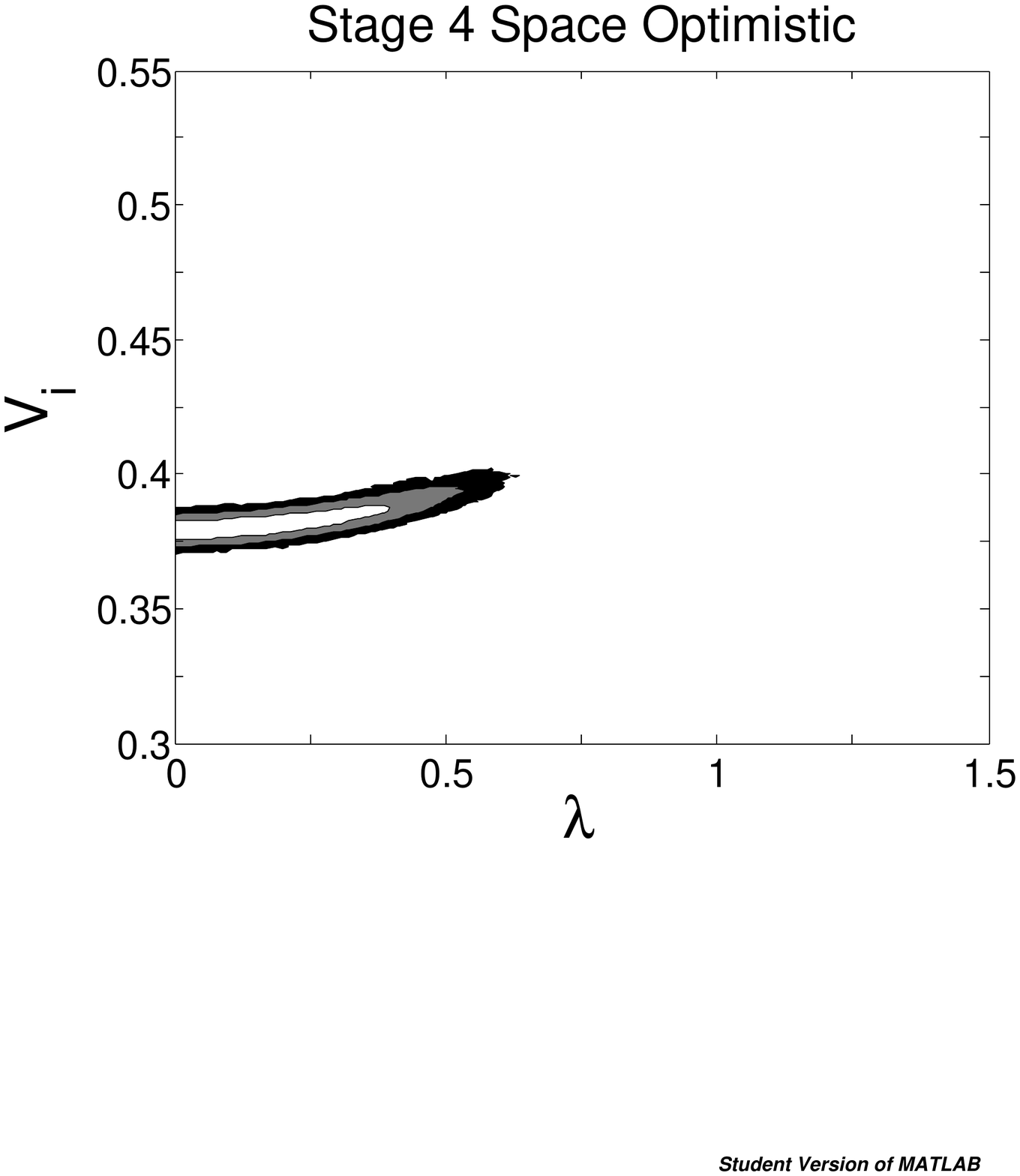}
\vspace{30pt}

\caption{Likelihood contours in the $V_I - \lambda$ space for cosmological
  constant data models.  The three
  contours give $68.27$\%, $95.44$\%, and $99.73$\% confidence regions.}

\label{fig:lambdafid-pars}

\end{figure*}

In order to run our MCMC analysis on the potential $V = V_0 e^{-\lambda\phi}$, we first need to make
a careful choice of parameters.  
The obvious choice of the
potential parameters $V_0$, $ \lambda$, and the initial field value $ \phi_I $ presents
several problems. Rewriting the potential $V = V_0 e^{-\lambda\phi}
\rightarrow V = e^{\ln(V_0)-\lambda\phi}$ reveals a degeneracy between
$\ln(V_0)$ and $\lambda \phi_I$. For fixed values of
$\lambda$, a change in $\phi_I$ and a corresponding change in
$\ln(V_0)$ gives identical cosmological solutions. This will lead
to an unconstrained and uninteresting parameter space. Fixing $V_0$
removes this degeneracy. We make the choice of $V_0 = \rho_\Lambda =
8.74\times10^{-121}$ (in reduced Planck units), which is the value of
the cosmological constant energy density used by the DETF. This is the
simplest choice, although not absolutely necessary.  Other choices of
$V_0$ would provide equivalent cosmological 
solutions.  

Removing the degeneracy and fixing $V_0$ leaves $\lambda$ and $\phi_I$ as the two
model parameters. However, this choice leads to an ``infinite direction''
in $\lambda-\phi_I$ space: Since the data for the first part of our
analysis is modeled on a cosmological constant, the most probable
values of $\lambda$ are those where $\lambda$ approaches zero. As $\lambda$
approaches zero, $\phi_I$ can take any value and
produce solutions indistinguishable from a cosmological constant. This
leads to an infinite unconstrained direction in parameter space that is 
uninteresting and also fatal to the MCMC techniques.

One can resolve this problem by placing a bound
on $\lambda$ or $\phi_I$. For small values
of $\lambda$, a bound placed on $\lambda$ is nearly equivalent to placing
a bound on $w(a=1)$. A choice of a bound on $\lambda$ can be chosen such
that the difference from $w = -1$ is small, however the choice is
arbitrary. Further, for data based on a $\Lambda$ universe, the closer
the bound is placed to $\lambda = 0$, the more the allowed region of
parameter space squeezes against this bound as smaller 
values of $\lambda$ allow a wider range of $\phi_I$, basically
partially restoring the degeneracy we are trying to eliminate. This
arbitrariness and distortion of allowed parameter ranges make bounding
$\lambda$ a poor choice for addressing the parameter space
degeneracies in this model. The squeezing effect leads to incorrect conclusions about allowed values
of $\lambda$.  The space appears to disfavor larger values of
$\lambda$, or equivalently larger departures from $w(a=1) = -1$, than
would the space in the final parameterization that we discuss next. 

We find the best choice of free model parameters to be $V_I$, where
$V_I = V(\phi_I)$, and $\lambda$.  The value of $\phi_I$ is then
determined from $\lambda$ and $V_I$. This parameterization avoids the
degeneracy discussed in the previous paragraph since a cosmological
constant of a particular value is only represented at one point
($V_I = \rho_\Lambda$ and $\lambda = 0$) in the $\lambda-V_I$ space. Values
similar to a cosmological constant are explored without an arbitrary
bound placed on any parameter. This allows the MCMC method freedom to
explore a more natural space.  

There is no loss of generality with this choice of parameterization as
is easily seen by the slope and curvature of the potential in the new
parameters: 
\begin{equation}
\frac{dV}{d\phi} = -\lambda V_I , \frac{d^2V}{d\phi^2} = \lambda^2 V_I
\end{equation}
This parameterization allows a simple intuition about the role of
these parameters forming the solutions.
Small values of $\lambda$ give a flat potential and the scalar field
will be stationary, independent of the choice of $V_I$. For
large values of $\lambda$ the field will roll and the amount to which
it does will depend on the value of $V_I$.

\begin{figure*}[!t]
\centering
%\vspace{-80pt}
\includegraphics[trim = 0mm 50mm 0mm 40mm, clip,width=0.45\textwidth]{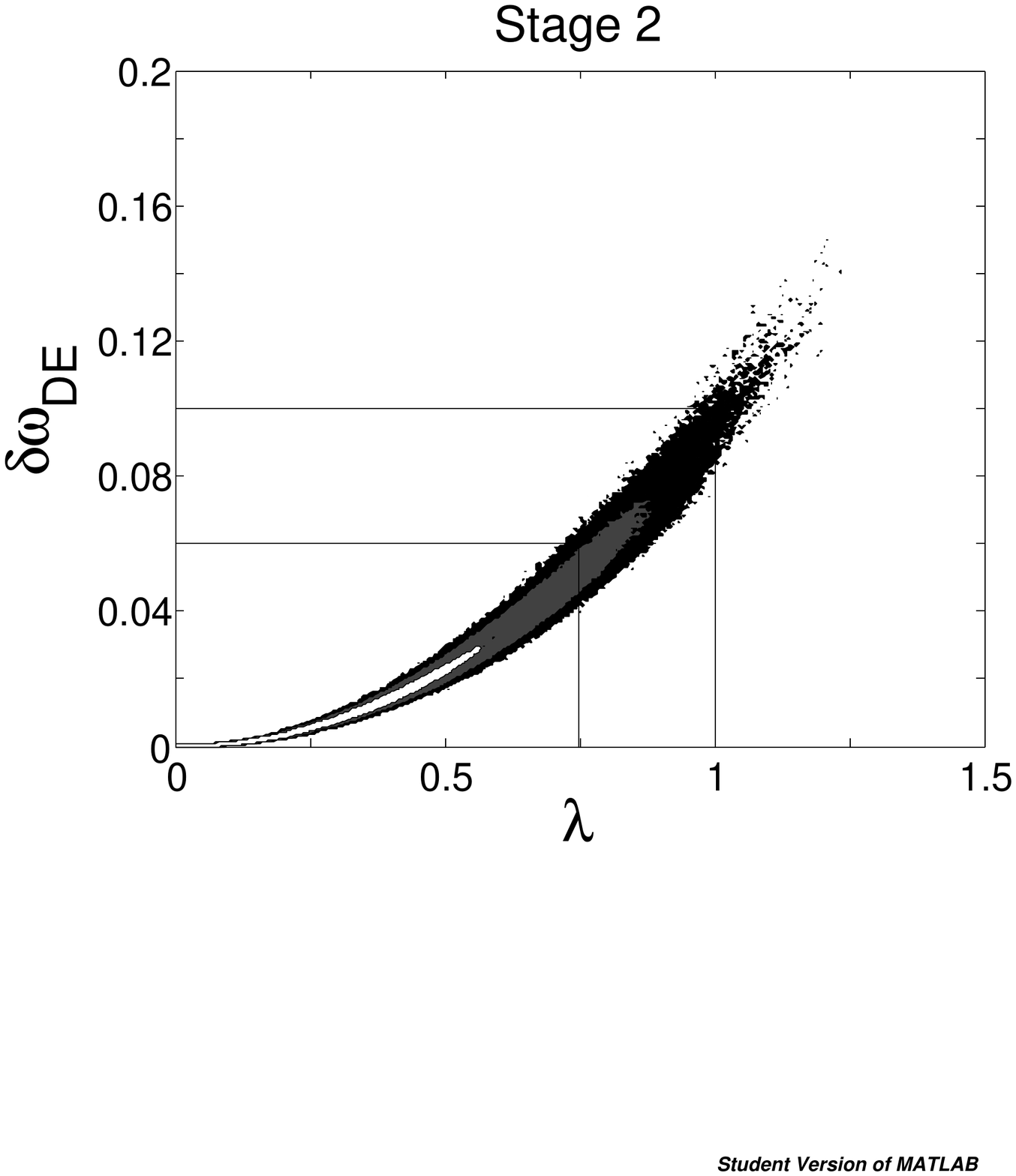}
\vspace{-20pt}
\includegraphics[trim = 0mm 50mm 0mm 40mm, clip,width=0.45\textwidth]{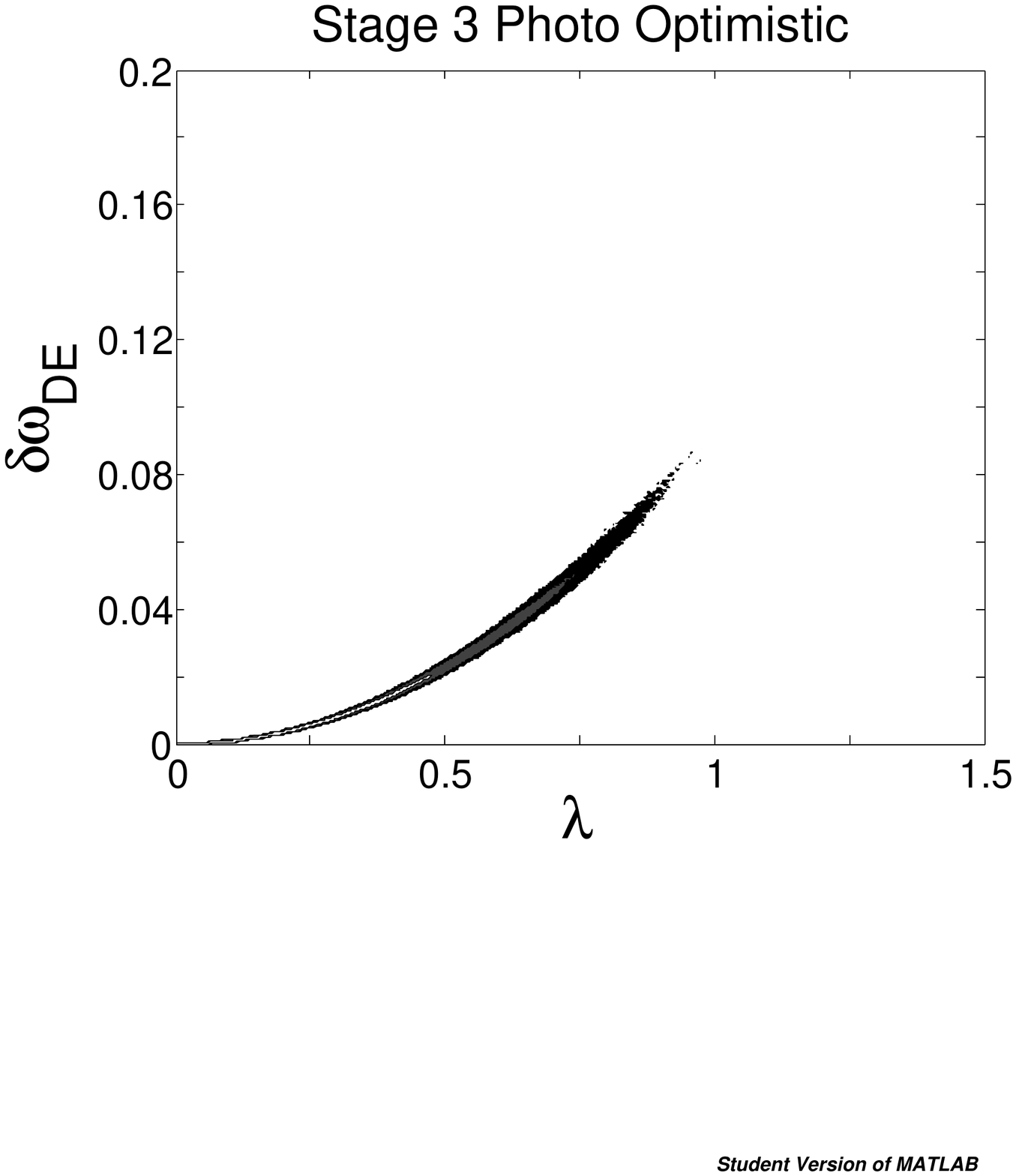}
\vspace{-20pt}
\includegraphics[trim = 0mm 50mm 0mm 40mm, clip,width=0.45\textwidth]{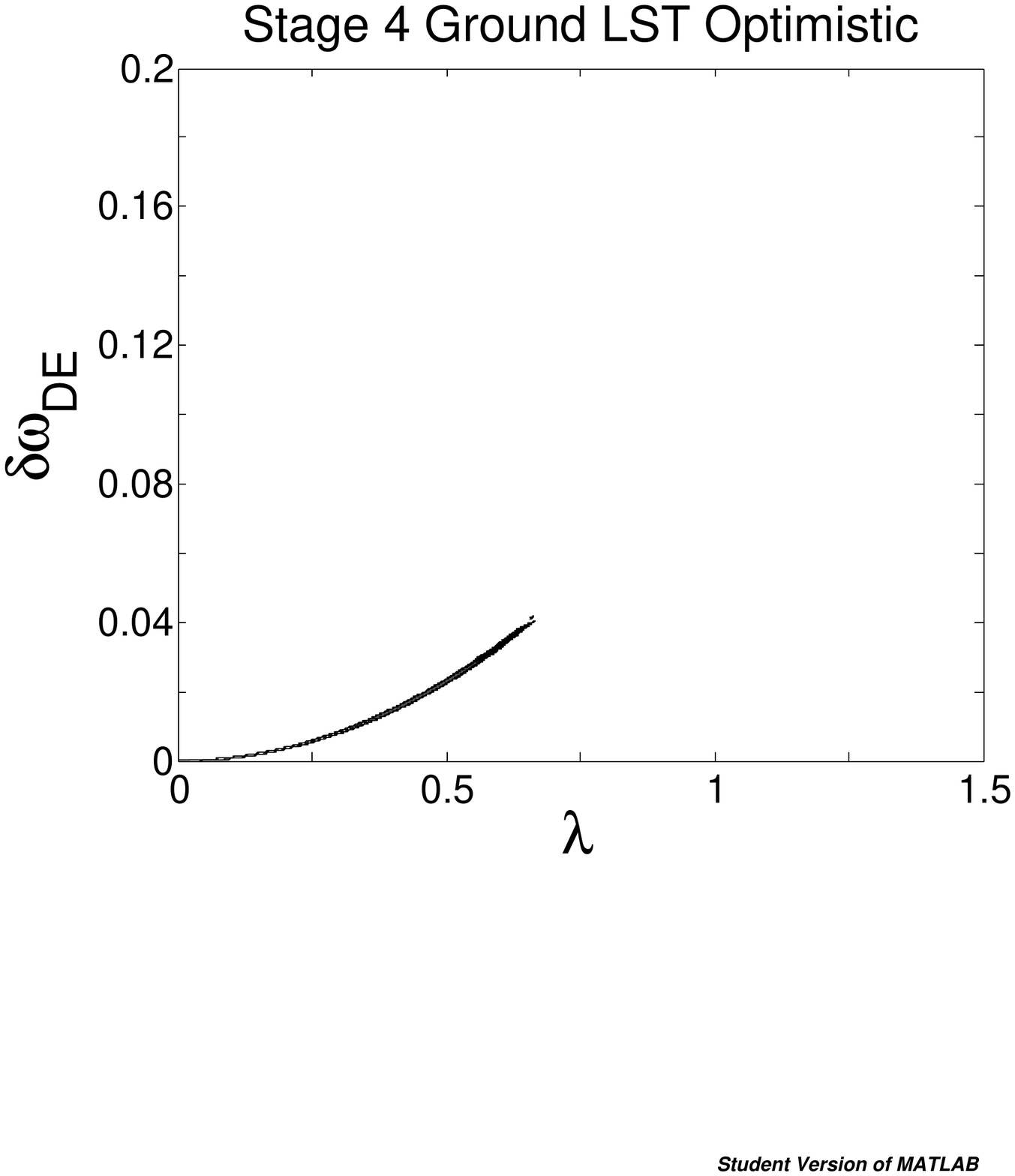}
\vspace{-20pt}
\includegraphics[trim = 0mm 50mm 0mm 40mm, clip,width=0.45\textwidth]{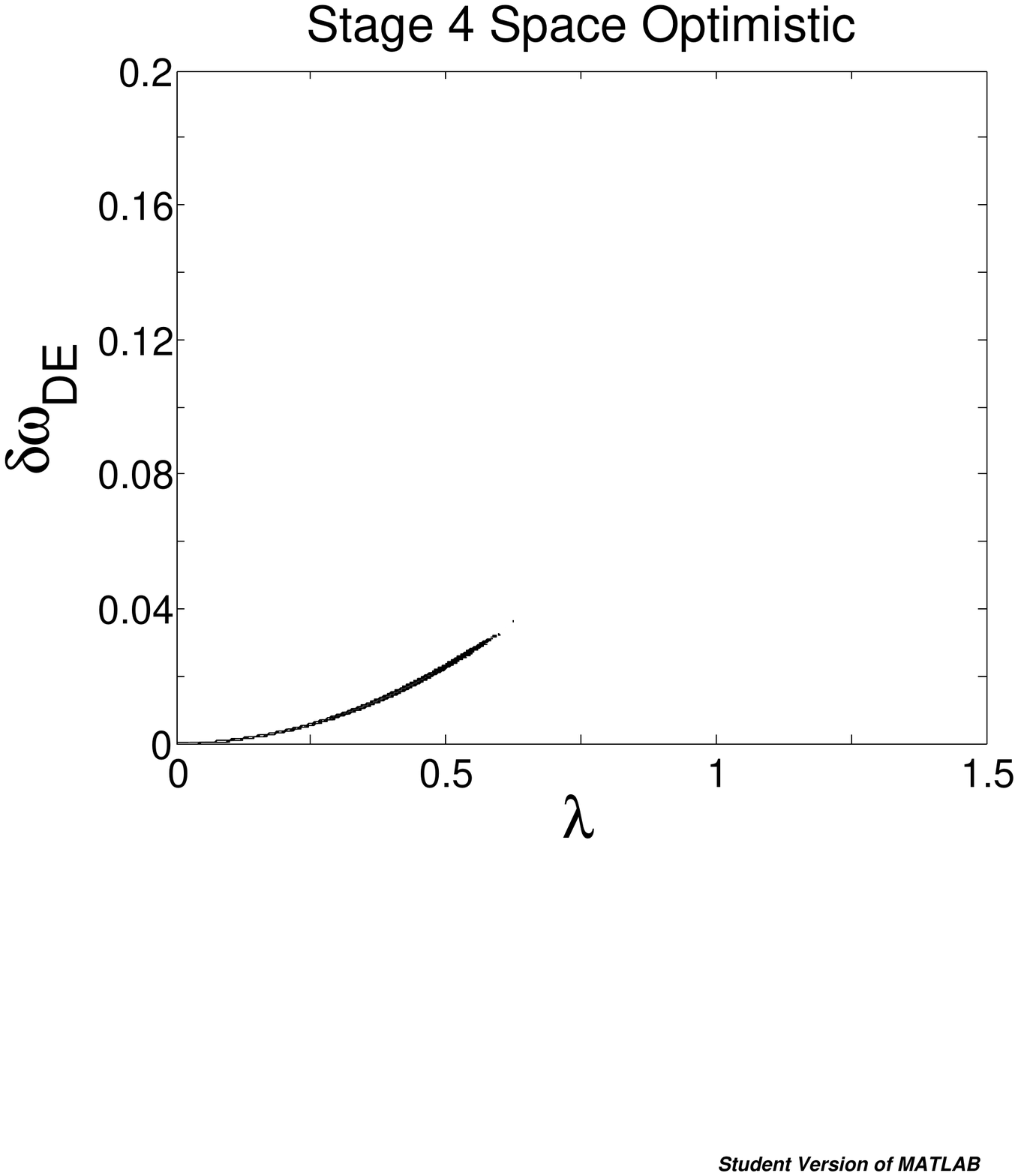}
\vspace{30pt}

\caption{Likelihood contours in the $\lambda-\delta\omega_{DE}$.
  Deviations of $\delta \omega_{DE}$ from $0$ indicate evolving dark
  energy.  The background cosmology has a cosmological constant.
The boxes in the top left panel
  (Stage 2) respectively show the size of the axes for Stage 3 and
  Stage 4 plots in Fig. \ref{fig:lamfid-delta-enlarge}. The
  contours give $68.27$\%, $95.44$\%, and $99.73$\% confidence regions.}  

\label{fig:lambdafid-delta}

\end{figure*}

\begin{figure*}[!t]
\centering
%\vspace{-80pt}
\includegraphics[trim = 0mm 50mm 0mm 40mm, clip,width=0.45\textwidth]{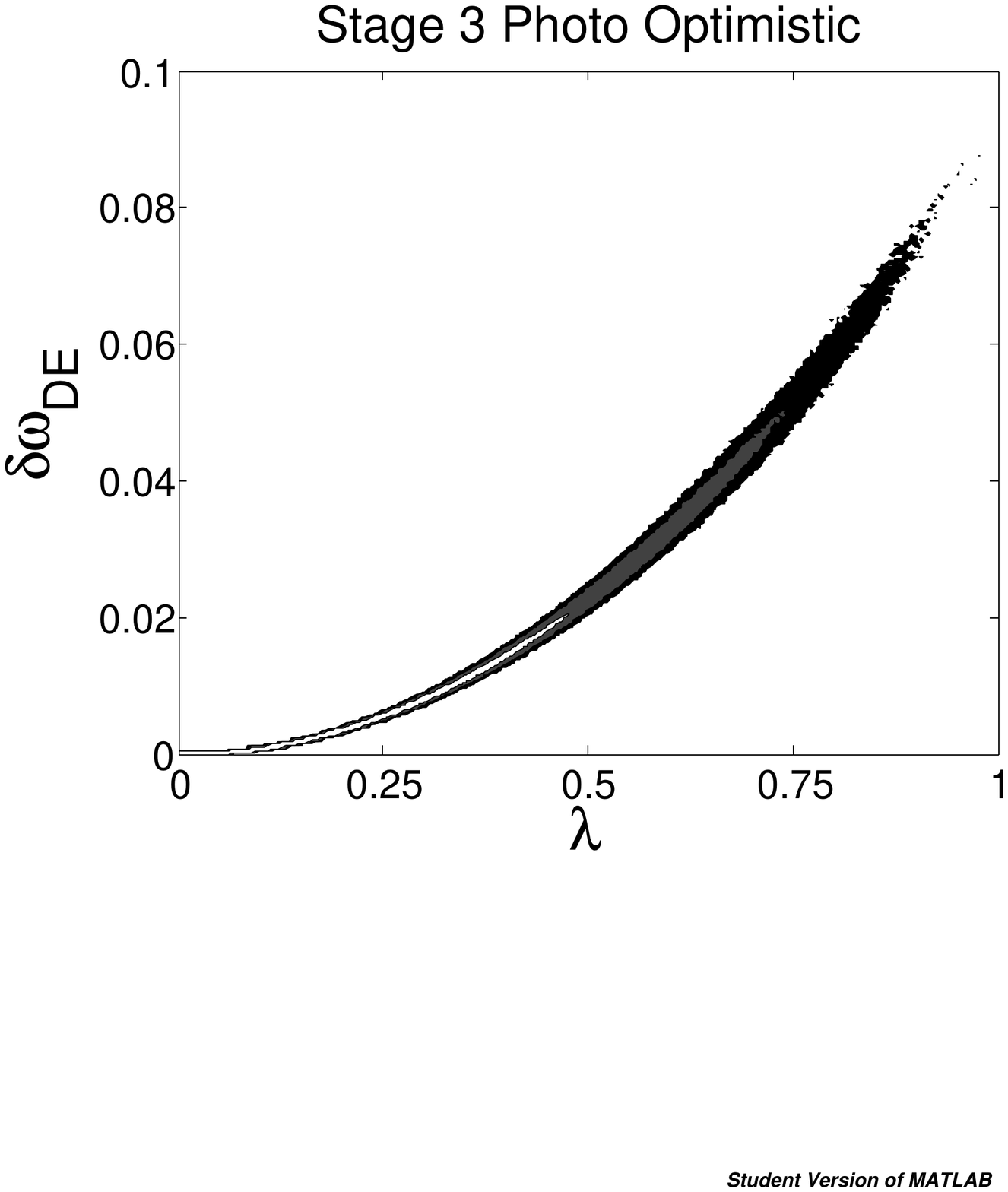}
\vspace{-20pt}
\includegraphics[trim = 0mm 50mm 0mm 40mm, clip,width=0.45\textwidth]{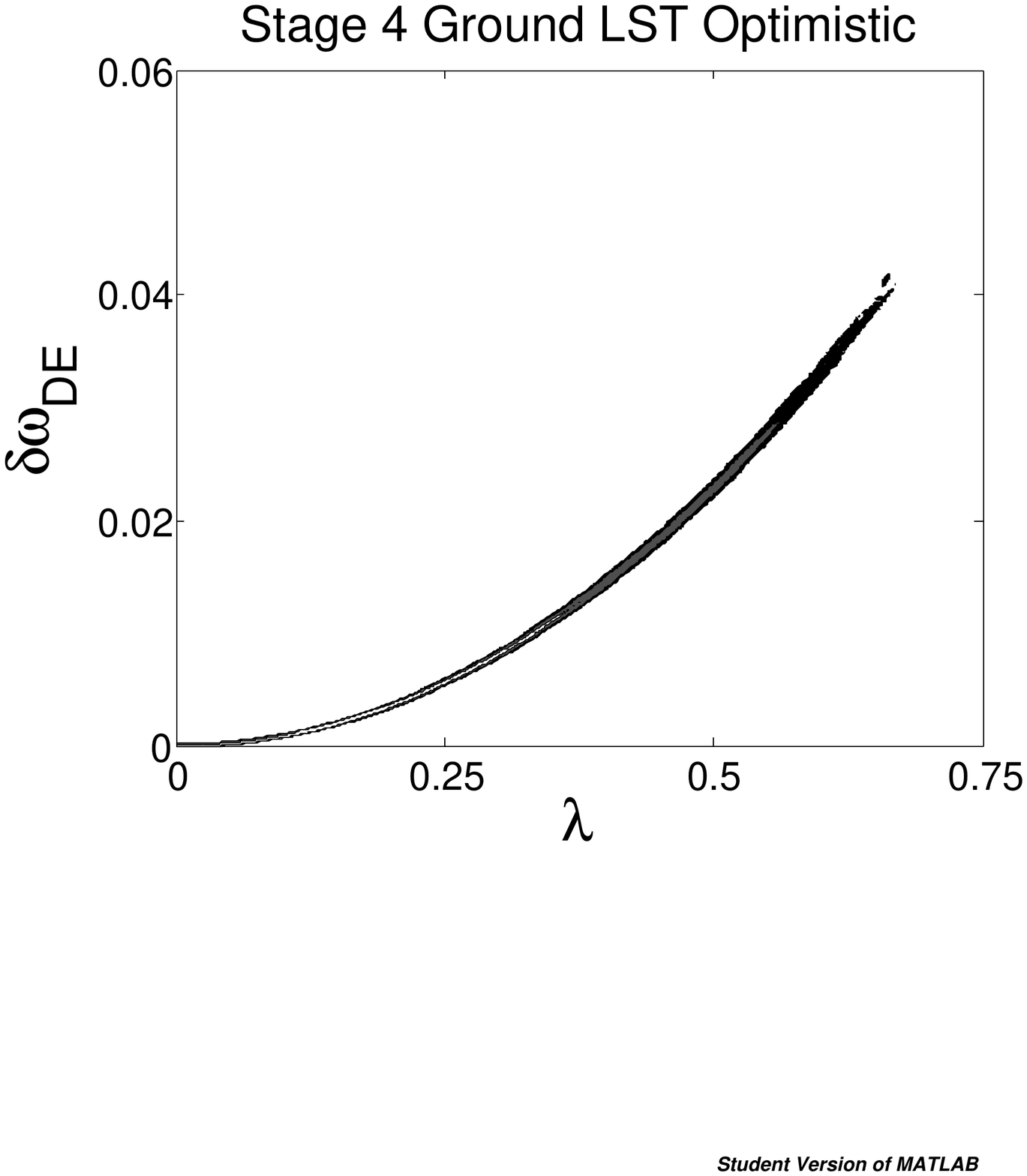}
\vspace{-20pt}
\includegraphics[trim = 0mm 50mm 0mm 40mm, clip,width=0.45\textwidth]{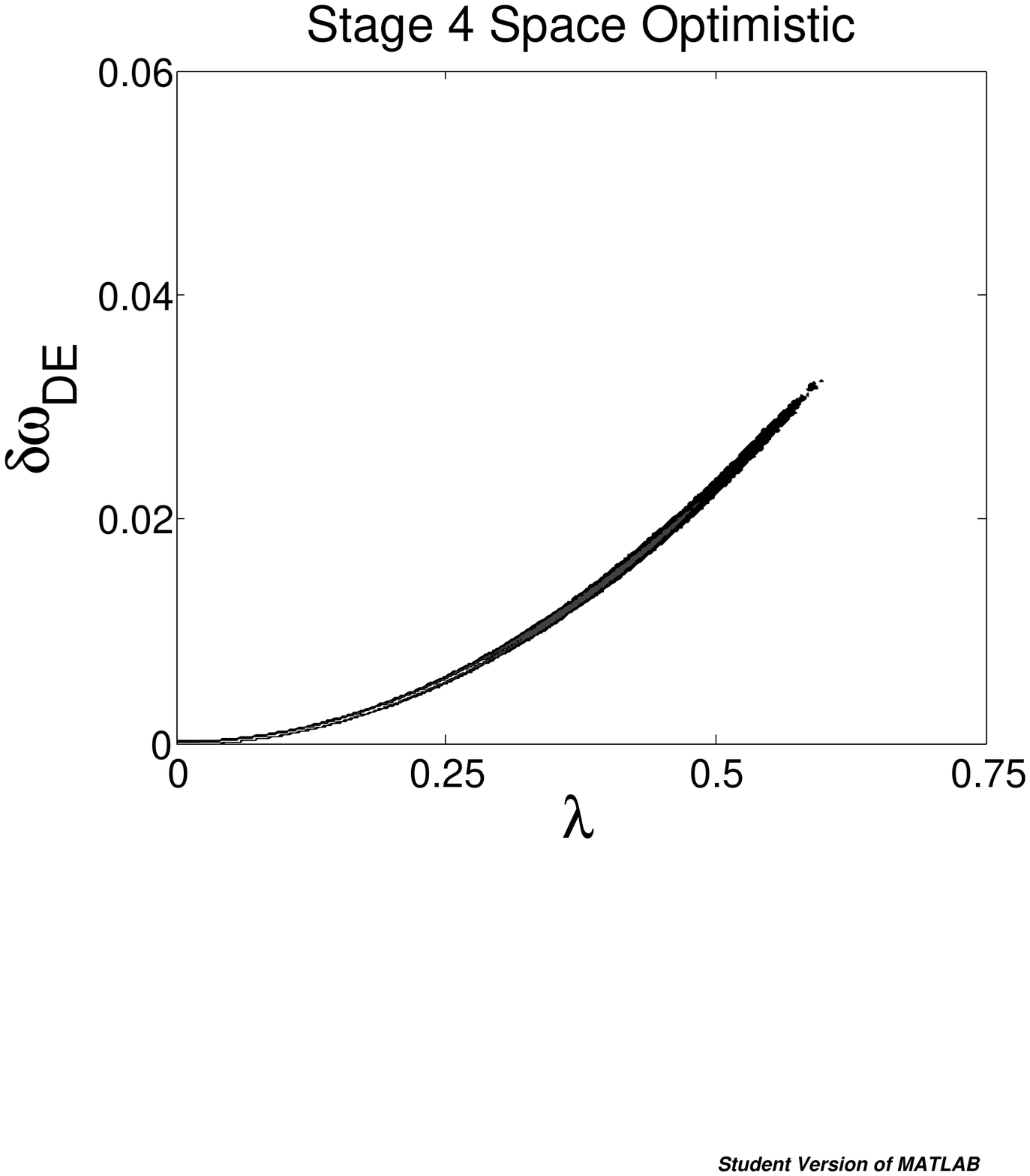}
\vspace{20pt}
\caption{Enlarged plots of the $\lambda-\delta\omega_{DE}$ confidence contours for
  cosmological constant data models. The three contours
  correspond to $68.27$\%,
  $95.44$\%, and $99.73$\% confidence regions. The axes correspond to the boxes in Fig. \ref{fig:lambdafid-delta}.} 
\label{fig:lamfid-delta-enlarge}

\end{figure*}

\begin{figure*}[!t]
\centering
%\vspace{-80pt}
\includegraphics[trim = 0mm 50mm 0mm 40mm, clip,width=0.45\textwidth]{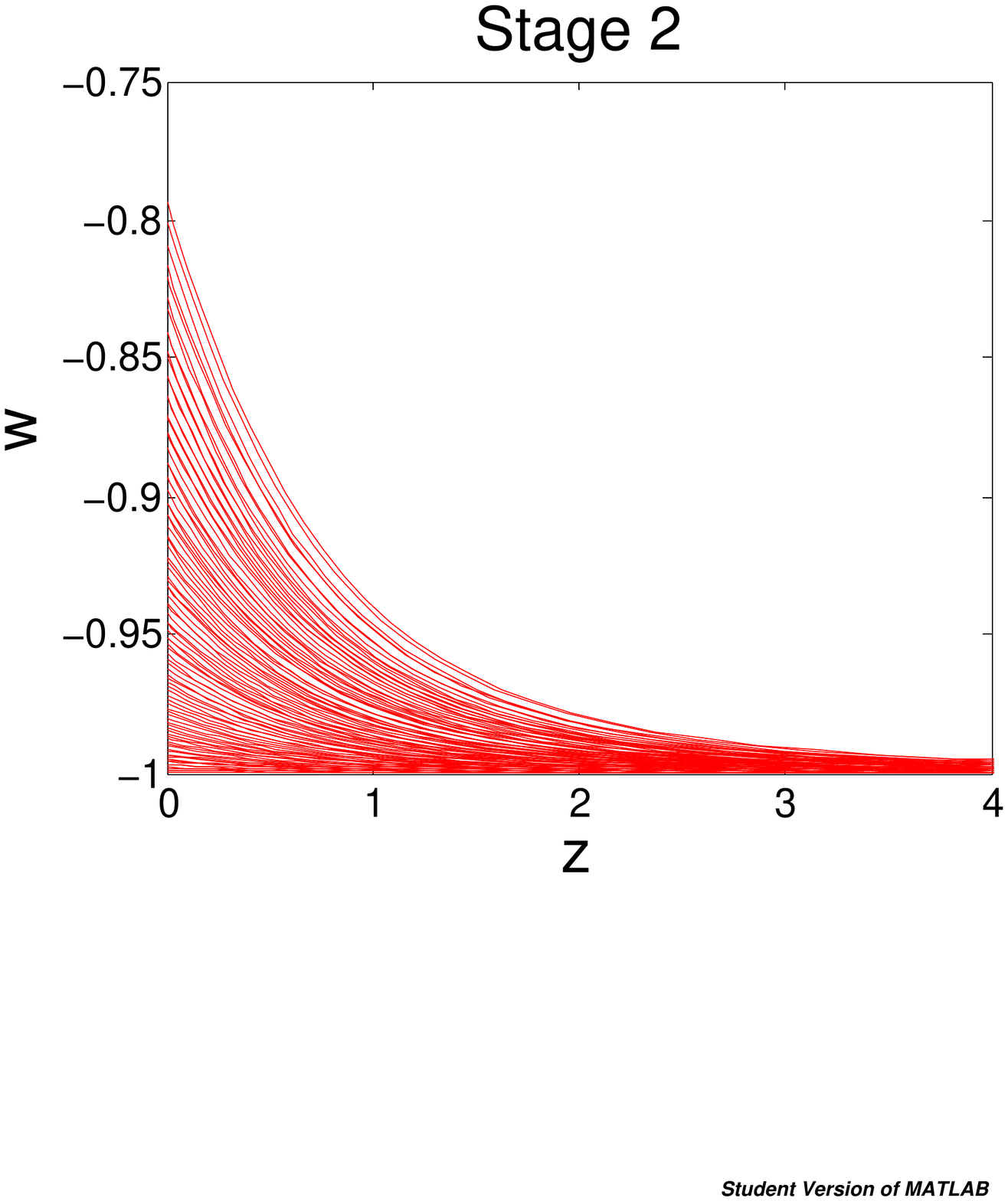}
\vspace{-20pt}
\includegraphics[trim = 0mm 50mm 0mm 40mm, clip,width=0.45\textwidth]{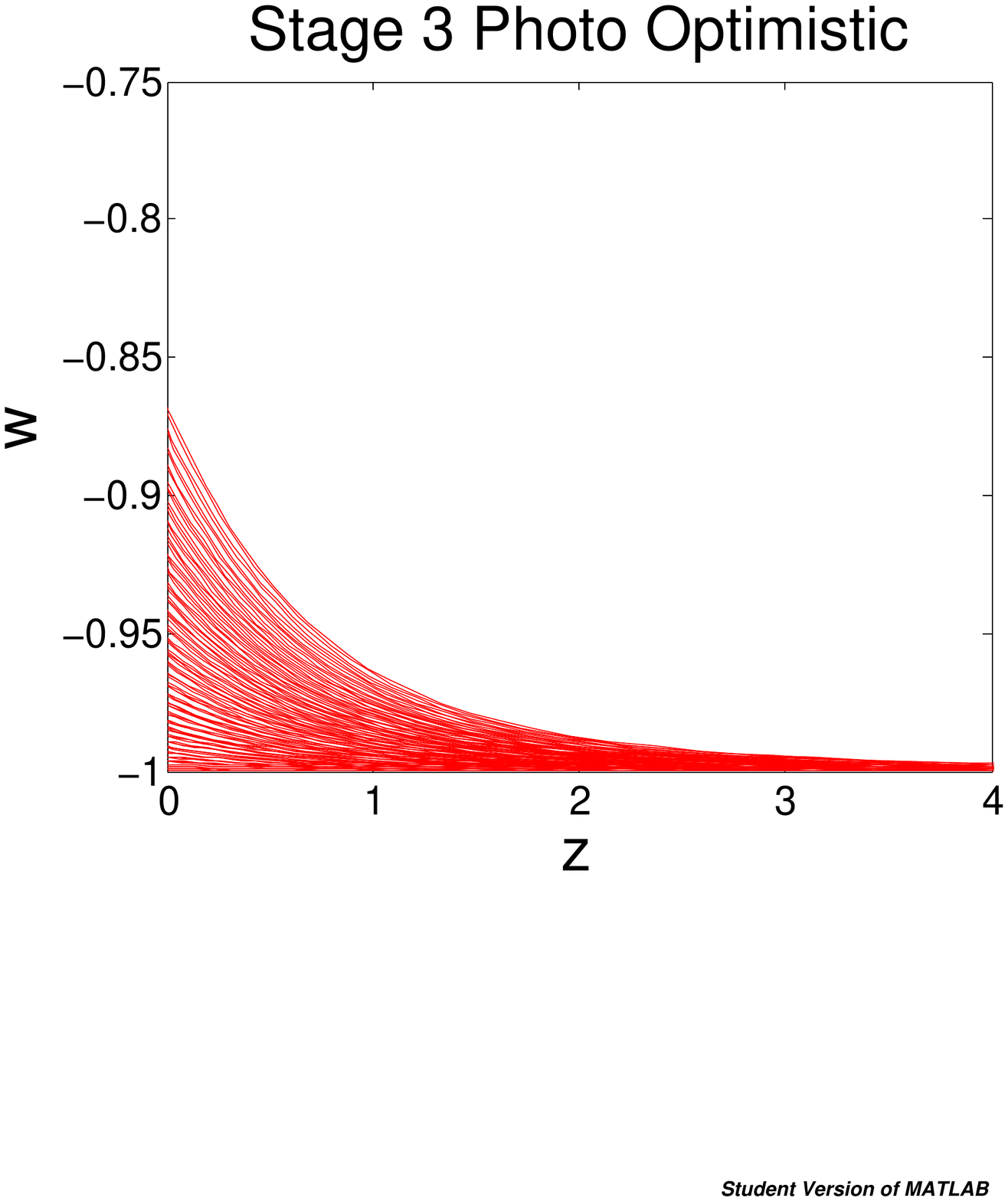}
\vspace{-20pt}
\includegraphics[trim = 0mm 50mm 0mm 40mm, clip,width=0.45\textwidth]{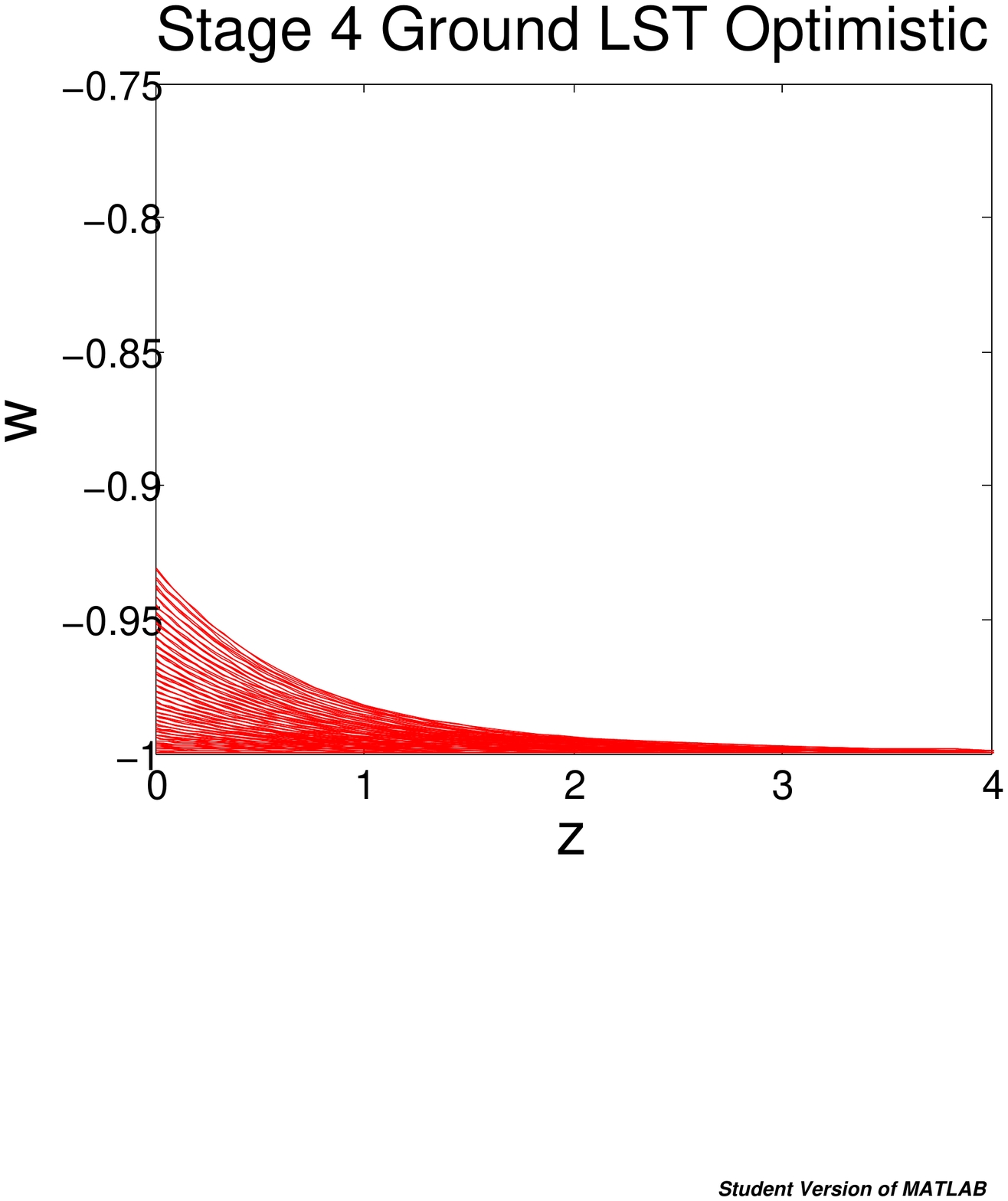}
\vspace{-20pt}
\includegraphics[trim = 0mm 50mm 0mm 40mm, clip,width=0.45\textwidth]{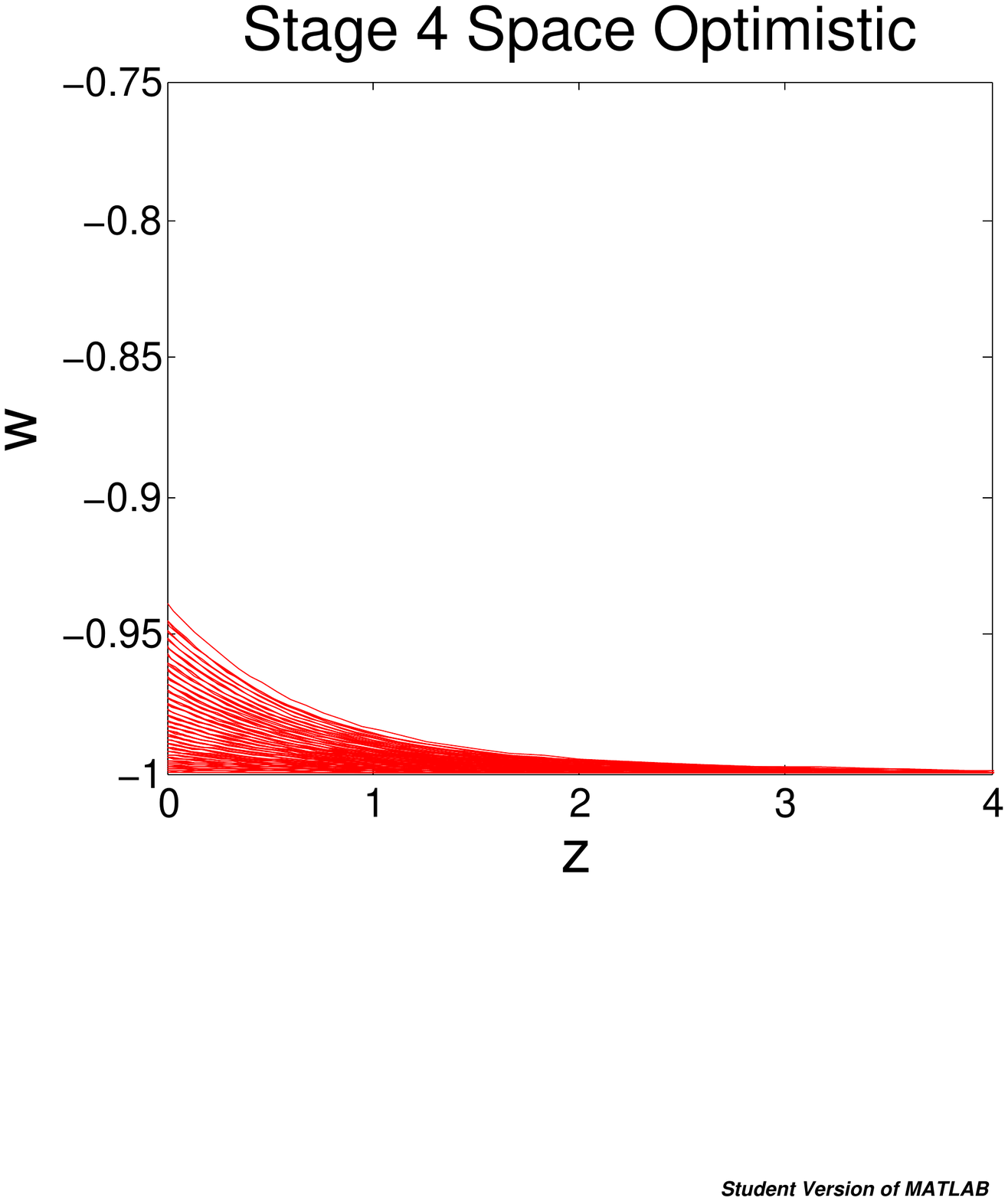}
\vspace{30pt}

\caption{The $w(z)$ behavior for a sample of points covering the full
  range of the $\lambda-V_I$ space for data based on a cosmological
  constant.} 
\label{fig:LambdaW(z)}

\end{figure*}

\section{\label{sec:Sec4}Cosmological Constant Fiducial Data}

\begin{figure*}[!t]
\centering
%\vspace{-80pt}
\includegraphics[trim = 0mm 50mm 0mm 40mm, clip,width=0.45\textwidth]{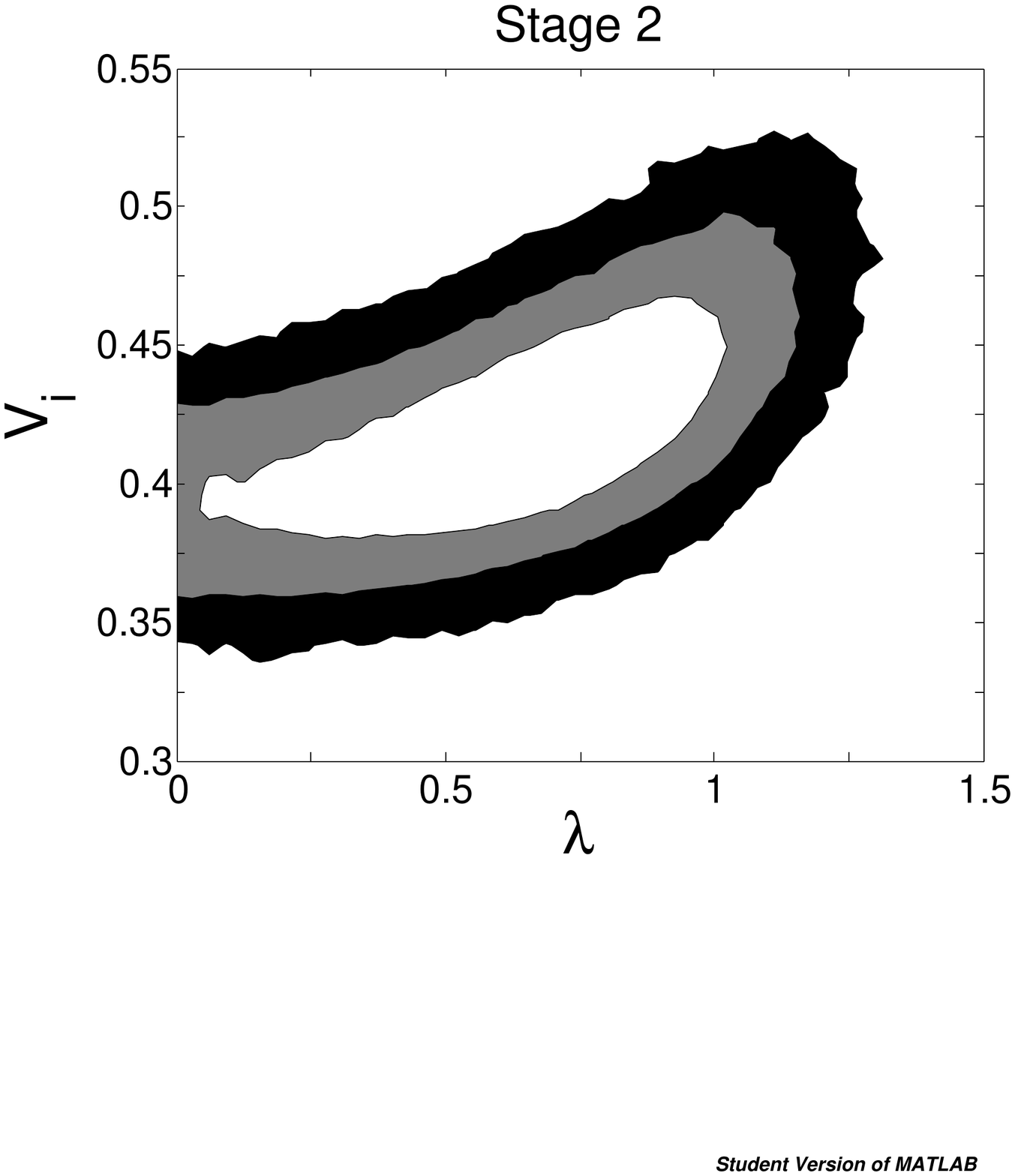}
\vspace{-20pt}
\includegraphics[trim = 0mm 50mm 0mm 40mm, clip,width=0.45\textwidth]{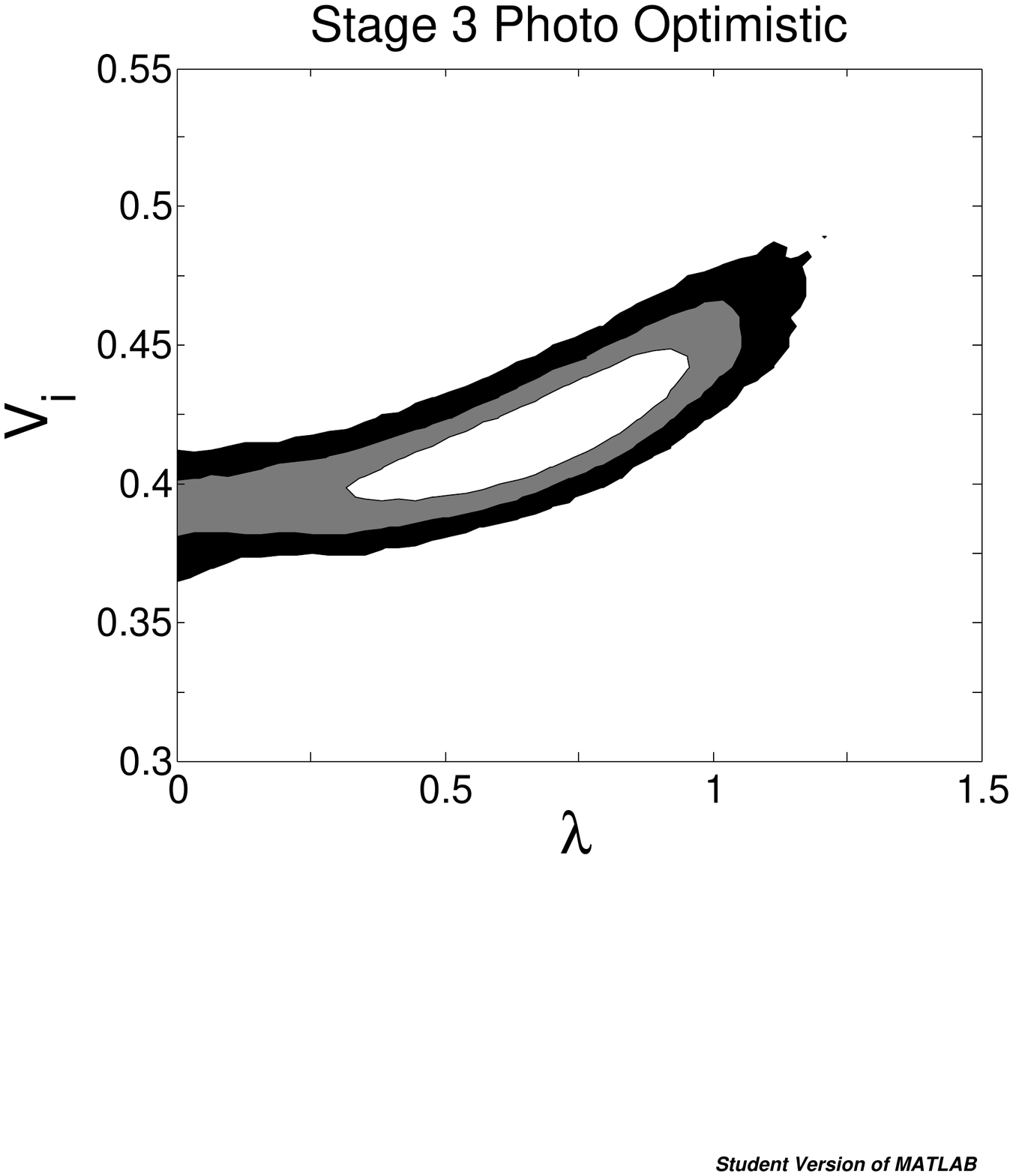}
\vspace{-20pt}
\includegraphics[trim = 0mm 50mm 0mm 40mm, clip,width=0.45\textwidth]{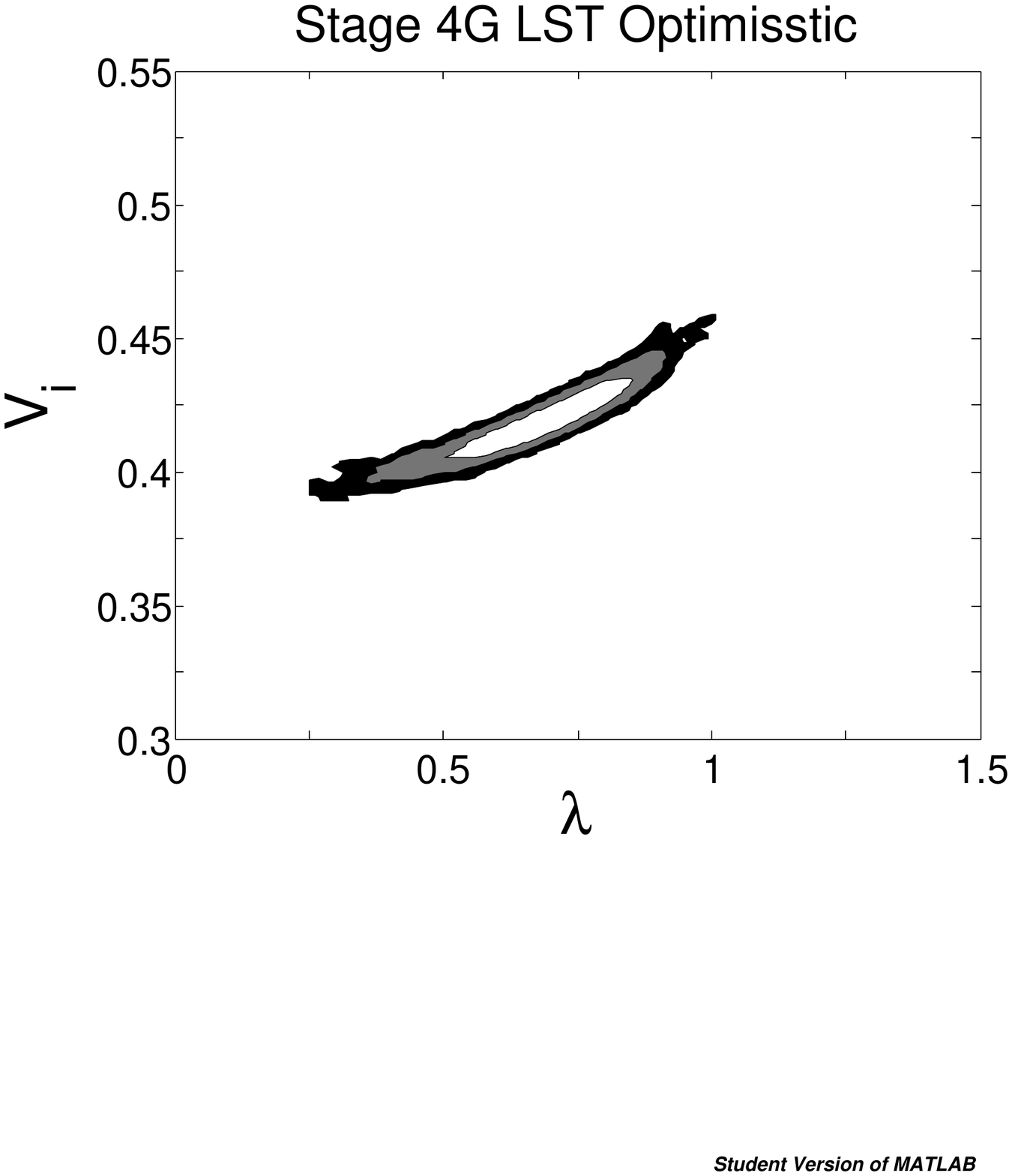}
\vspace{-20pt}
\includegraphics[trim = 0mm 50mm 0mm 40mm, clip,width=0.45\textwidth]{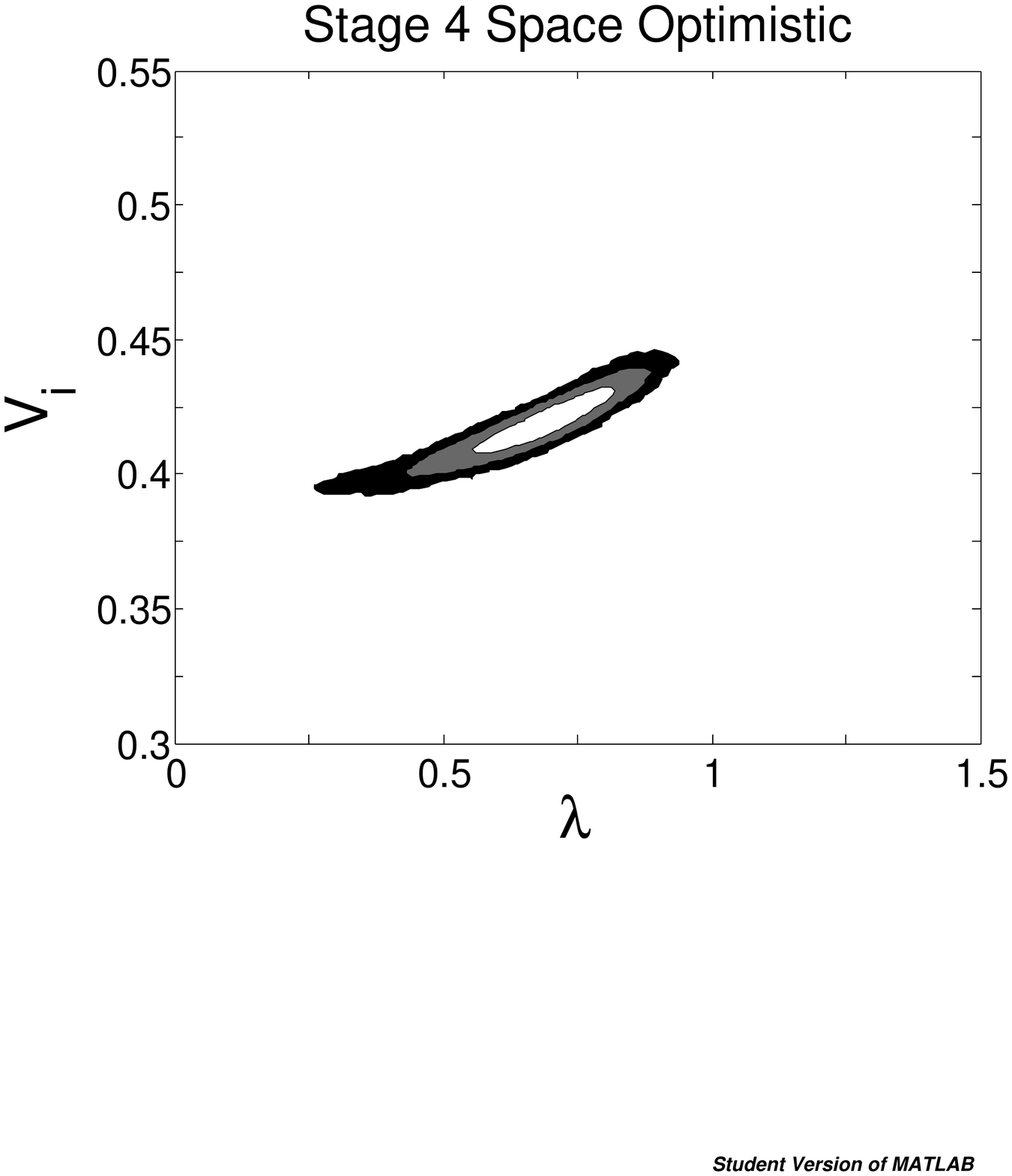}
\vspace{30pt}
\caption{Plots of the likelihood in $V_I - \lambda$ space for data
  based on the fiducial exponential background cosmology. The three contours
  correspond to $68.27$\%,
  $95.44$\%, and $99.73$\% confidence regions.}
\label{fig:Expfid-pars}

\end{figure*}

\begin{figure*}[!t]
\centering
%\vspace{-80pt}
\includegraphics[trim = 0mm 50mm 0mm 40mm, clip,width=0.45\textwidth]{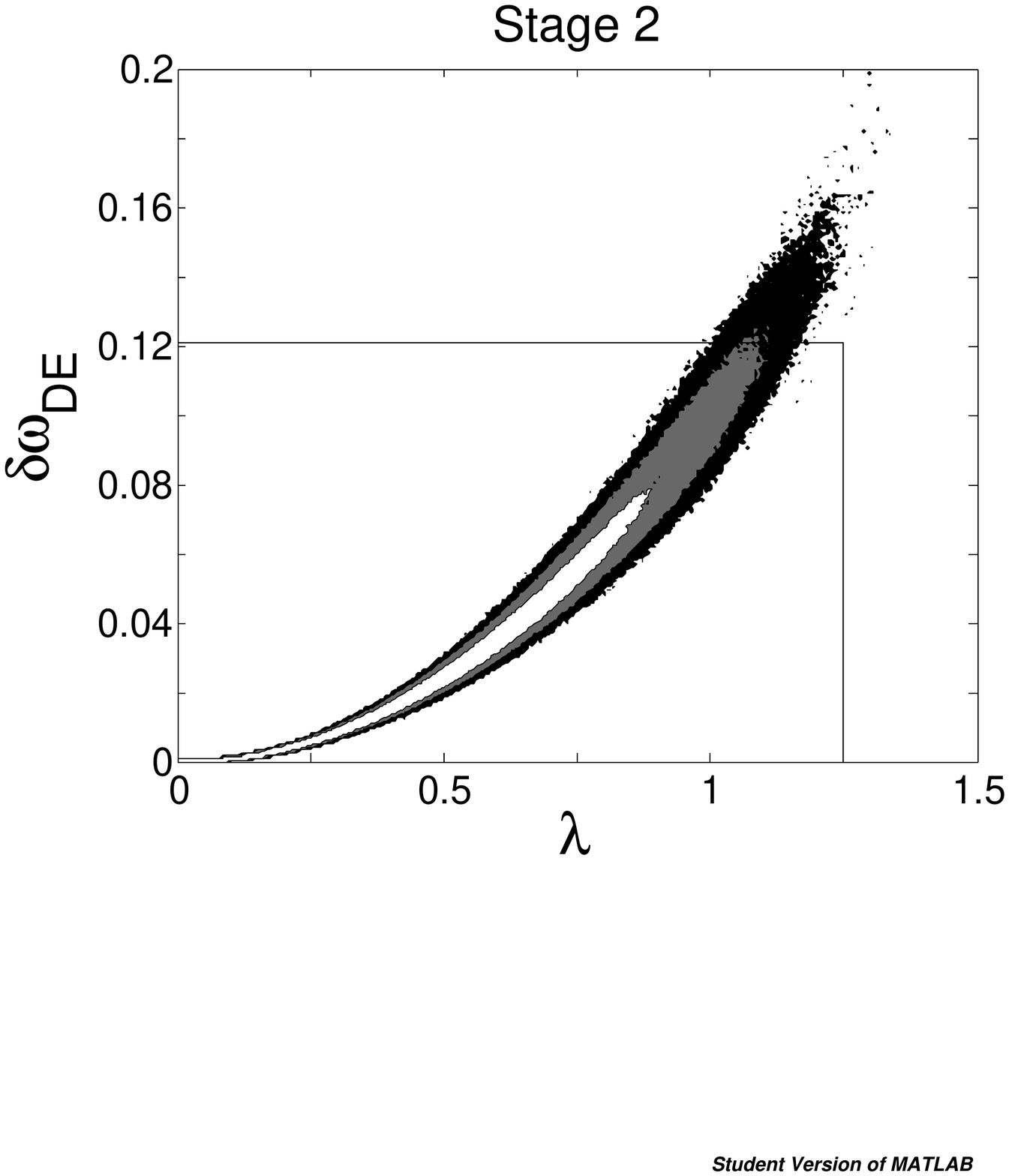}
\vspace{-20pt}
\includegraphics[trim = 0mm 50mm 0mm 40mm, clip,width=0.45\textwidth]{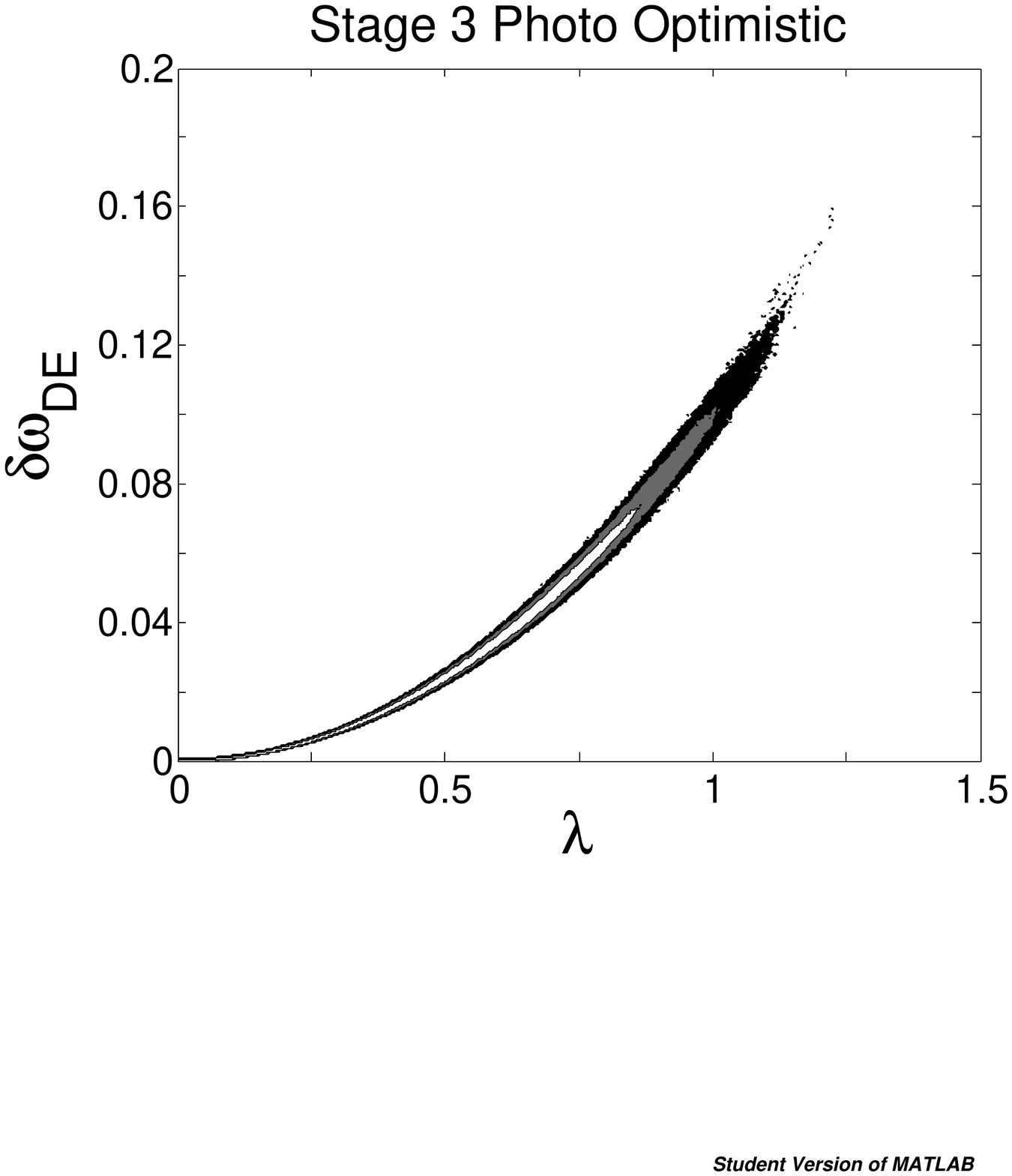}
\vspace{-20pt}
\includegraphics[trim = 0mm 50mm 0mm 40mm, clip,width=0.45\textwidth]{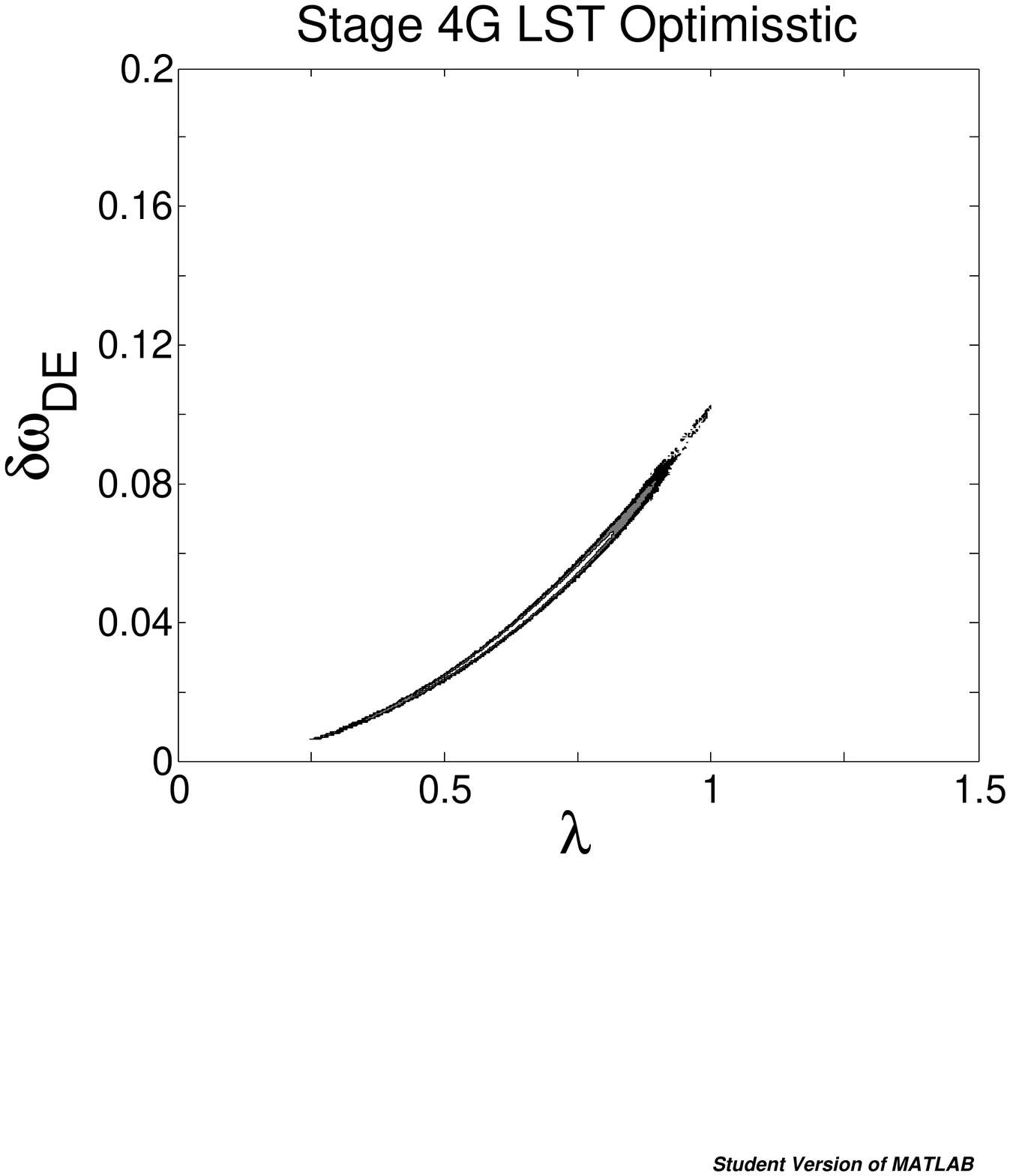}
\vspace{-20pt}
\includegraphics[trim = 0mm 50mm 0mm 40mm, clip,width=0.45\textwidth]{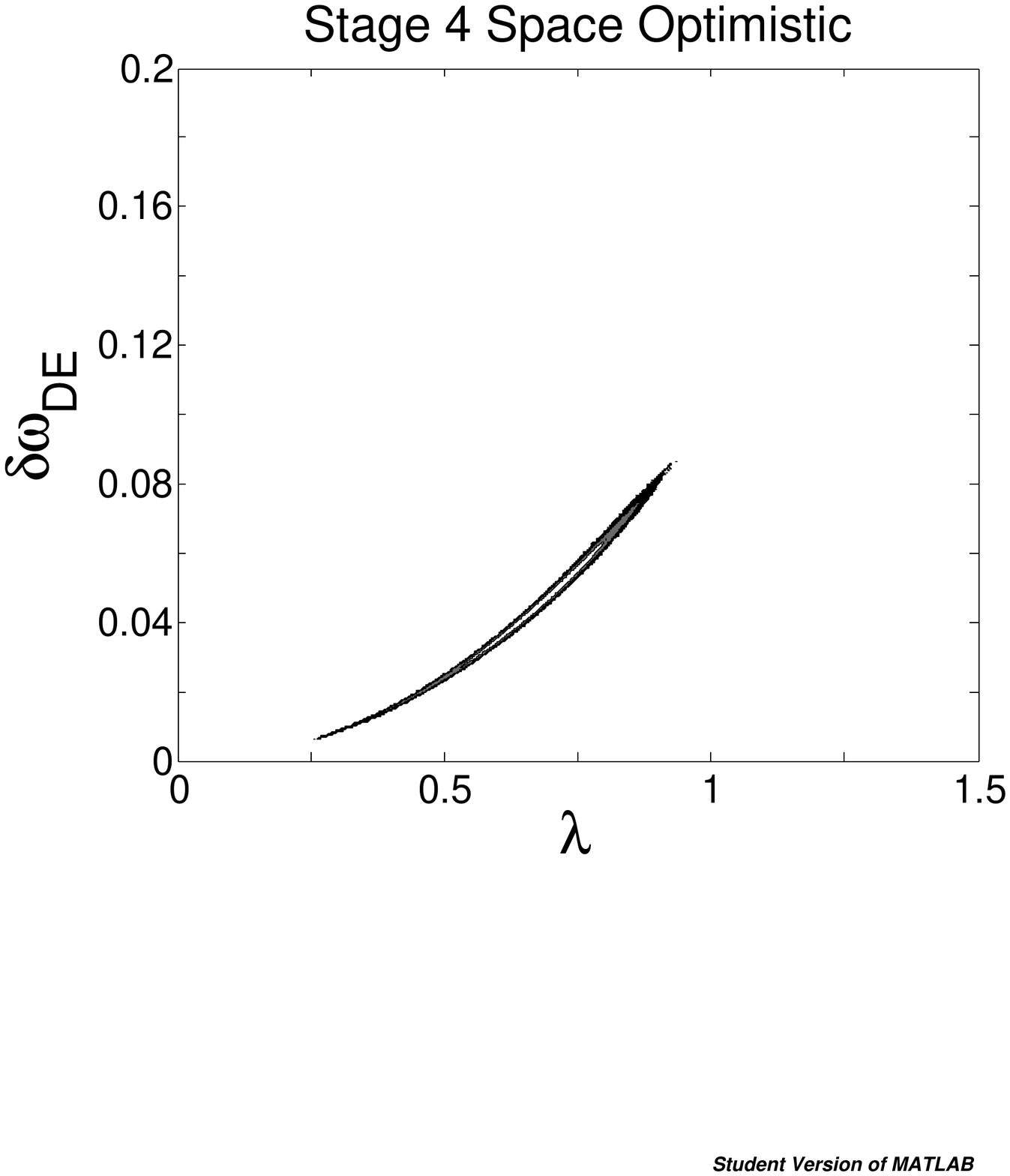}
\vspace{30pt}
\caption{Plots of the $\lambda-\delta\omega_{DE}$ confidence contours for data
  based on the fiducial exponential background cosmology.  The three contours
  correspond to $68.27$\%,
  $95.44$\%, and $99.73$\% confidence regions. The box in the top left figure
  (Stage 2) shows the size of the axes for the Stage 4 plots in Fig. \ref{fig:Expfid-delta-enlarge}.} 
\label{fig:Expfid-delta}

\end{figure*}

\begin{figure*}[!t]
\centering
%\vspace{-80pt}
\includegraphics[trim = 0mm 50mm 0mm 40mm, clip,width=0.45\textwidth]{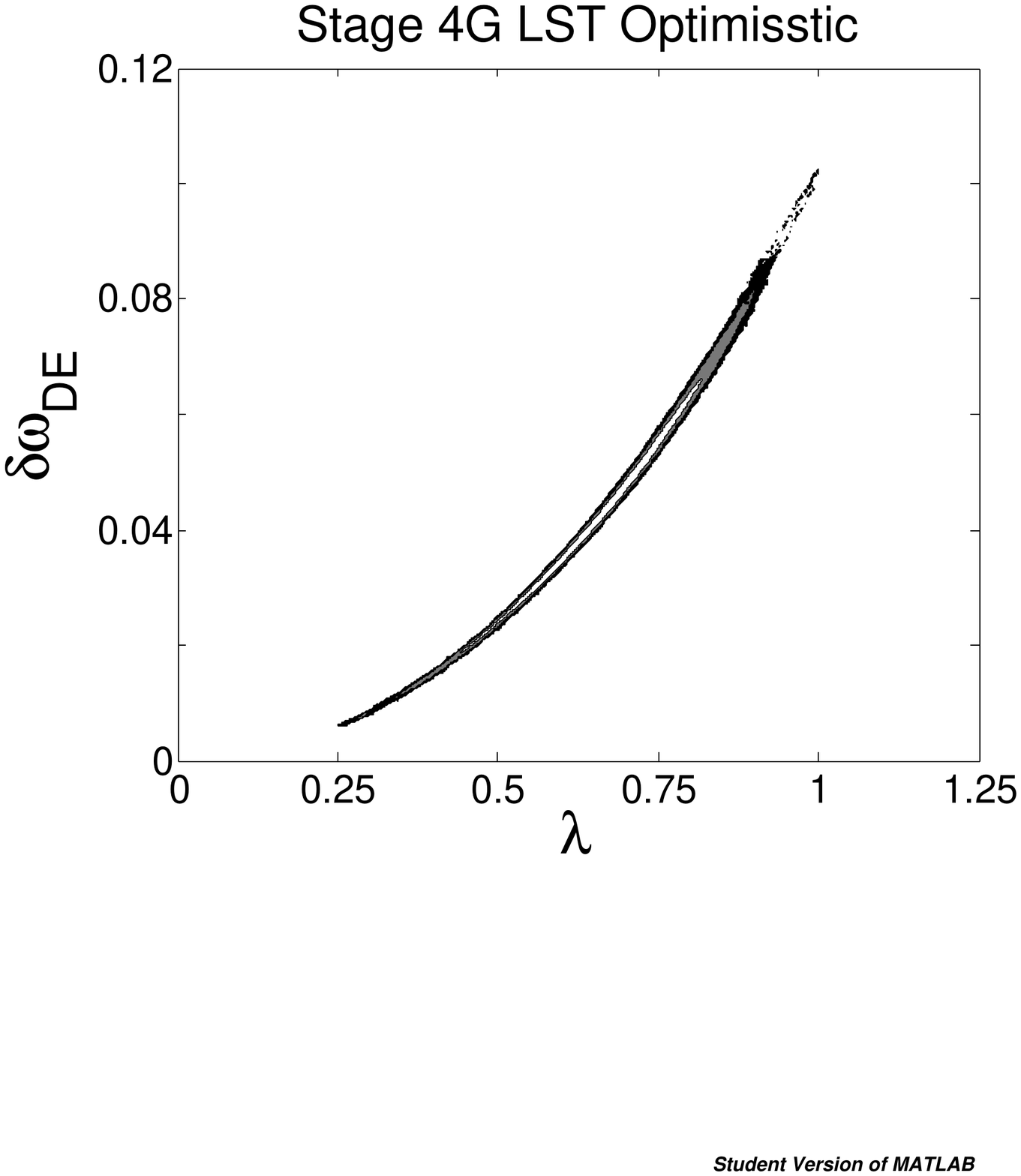}
\vspace{-20pt}
\includegraphics[trim = 0mm 50mm 0mm 40mm, clip,width=0.45\textwidth]{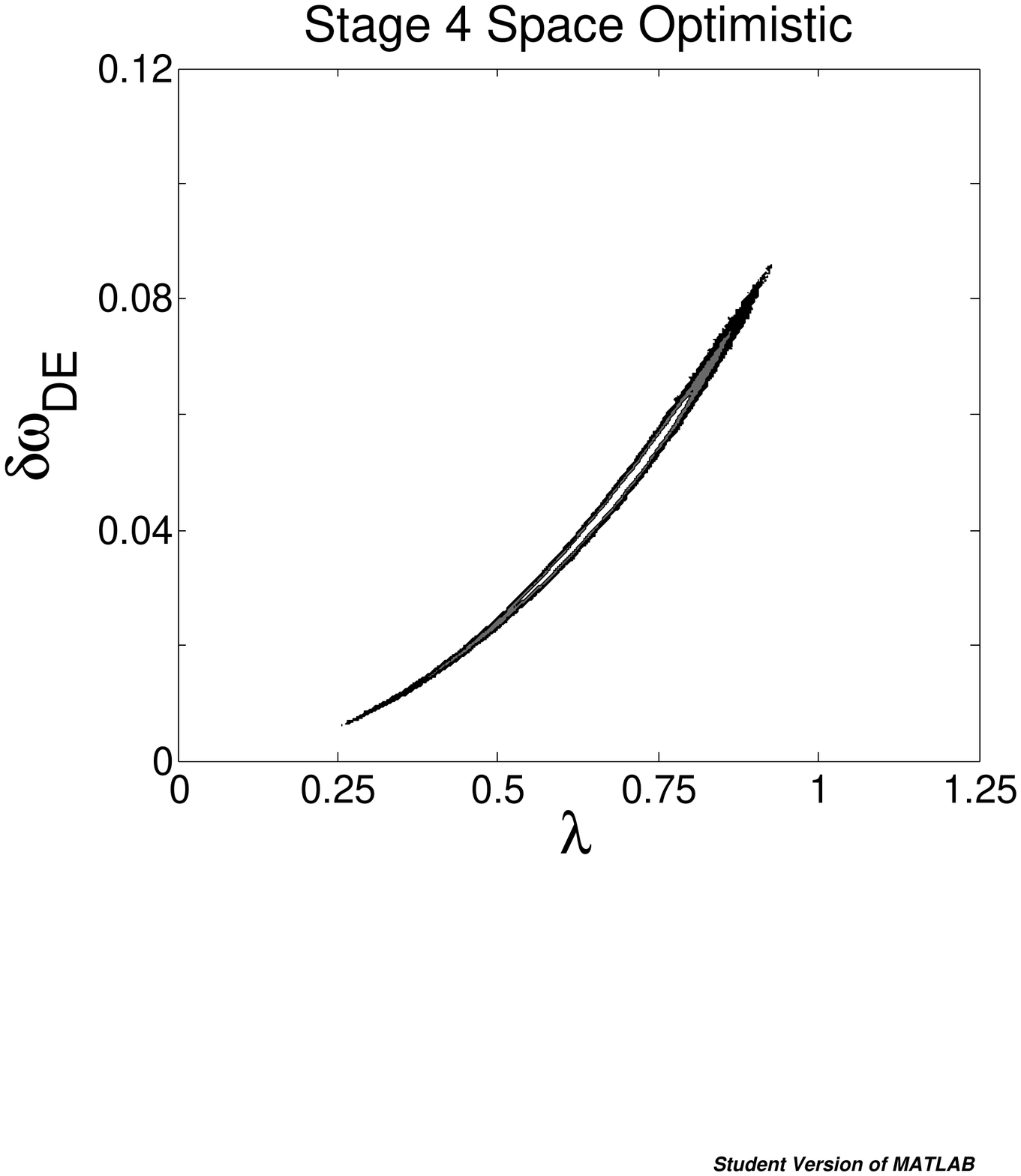}
\vspace{30pt}
\caption{Enlarged plots of the $\lambda-\delta\omega_{DE}$ confidence contours for data
  based on an exponential model.  The three contours
  correspond to $68.27$\%,
  $95.44$\%, and $99.73$\% confidence regions. The axes correspond to the boxes in Fig. \ref{fig:Expfid-delta}.} 
\label{fig:Expfid-delta-enlarge}

\end{figure*}

\begin{figure*}[!t]
\centering
%\vspace{-80pt}
\includegraphics[trim = 0mm 50mm 0mm 40mm, clip,width=0.45\textwidth]{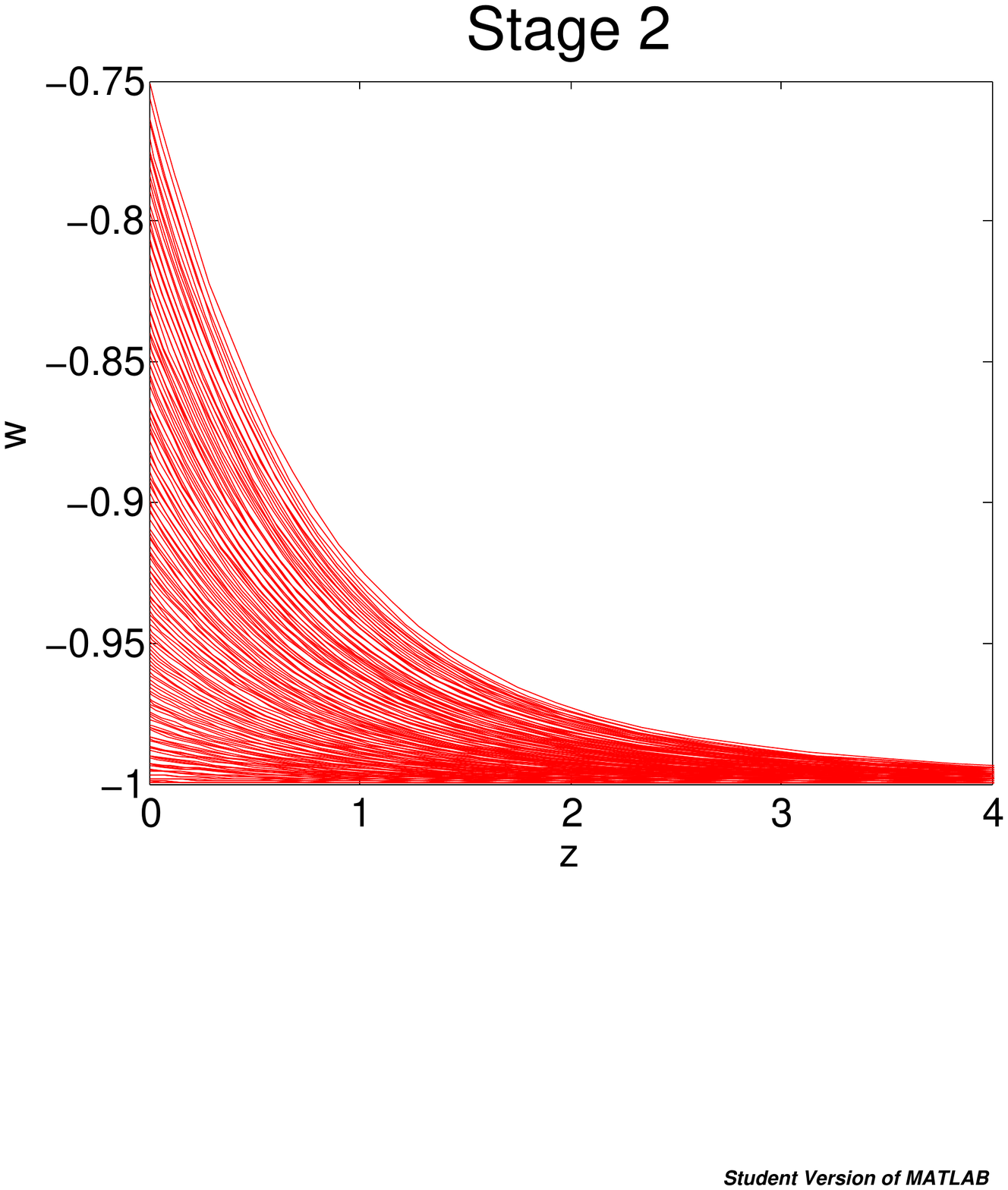}
\vspace{-20pt}
\includegraphics[trim = 0mm 50mm 0mm 40mm, clip,width=0.45\textwidth]{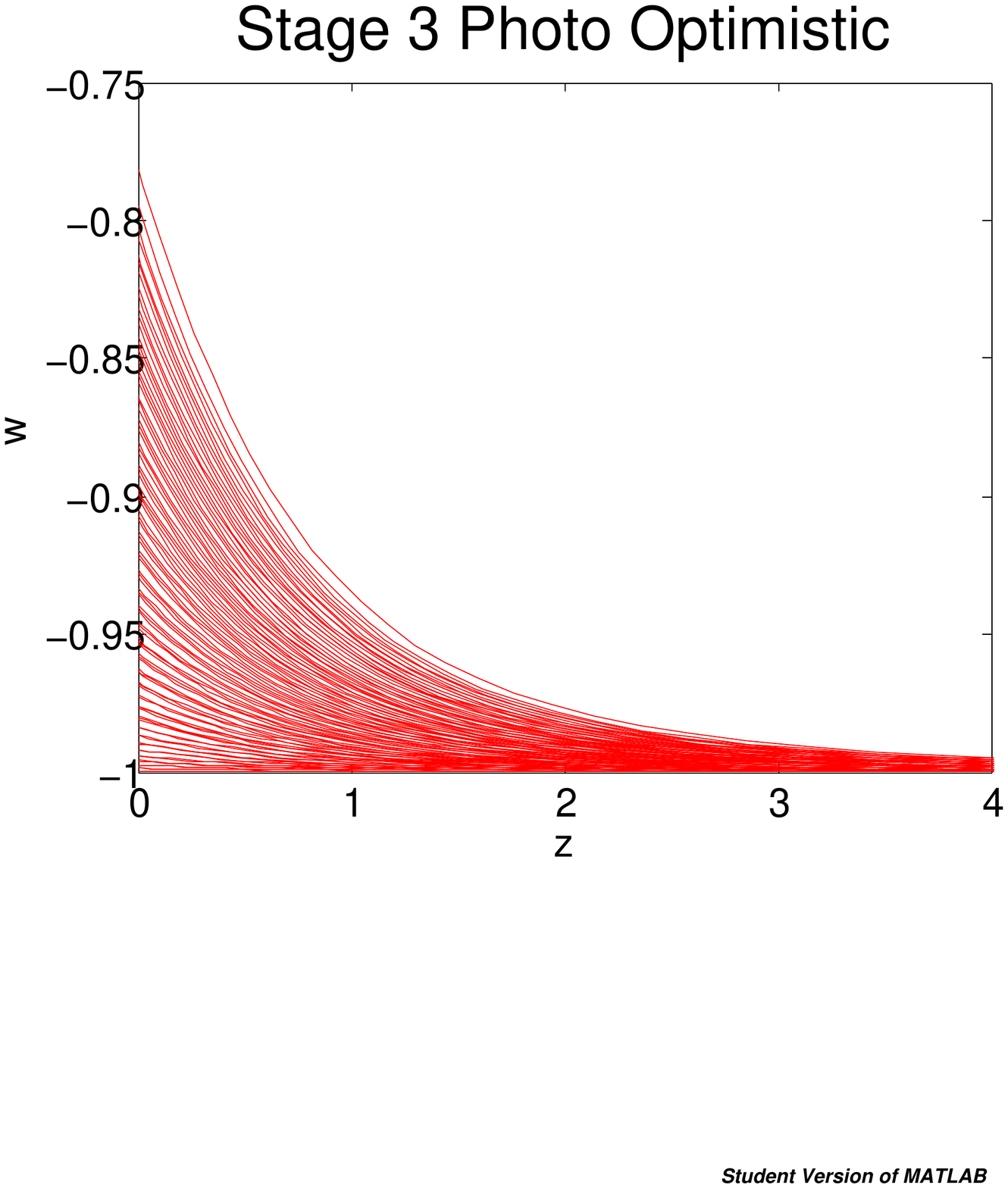}
\vspace{-20pt}
\includegraphics[trim = 0mm 50mm 0mm 40mm, clip,width=0.45\textwidth]{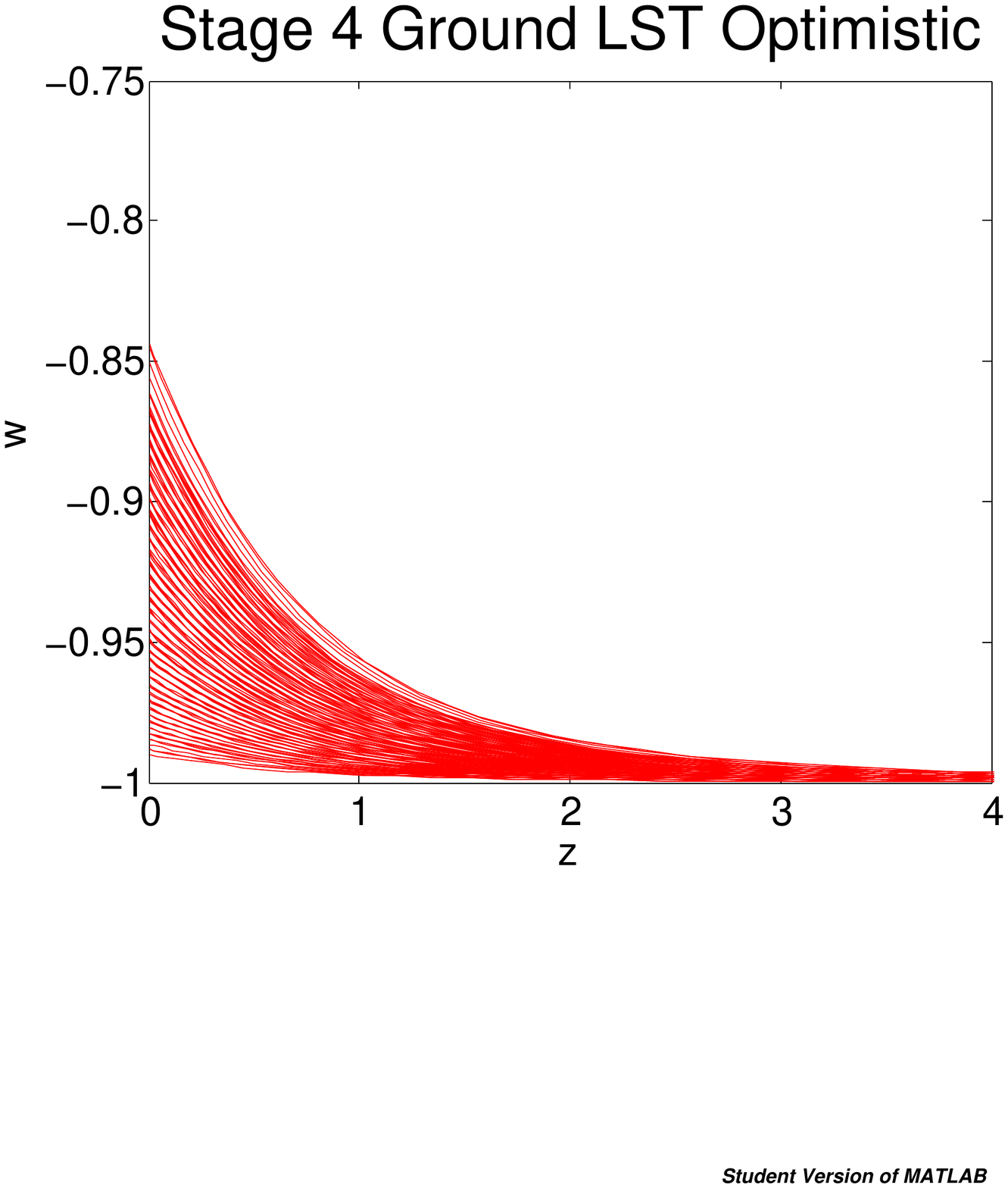}
\vspace{-20pt}
\includegraphics[trim = 0mm 50mm 0mm 40mm, clip,width=0.45\textwidth]{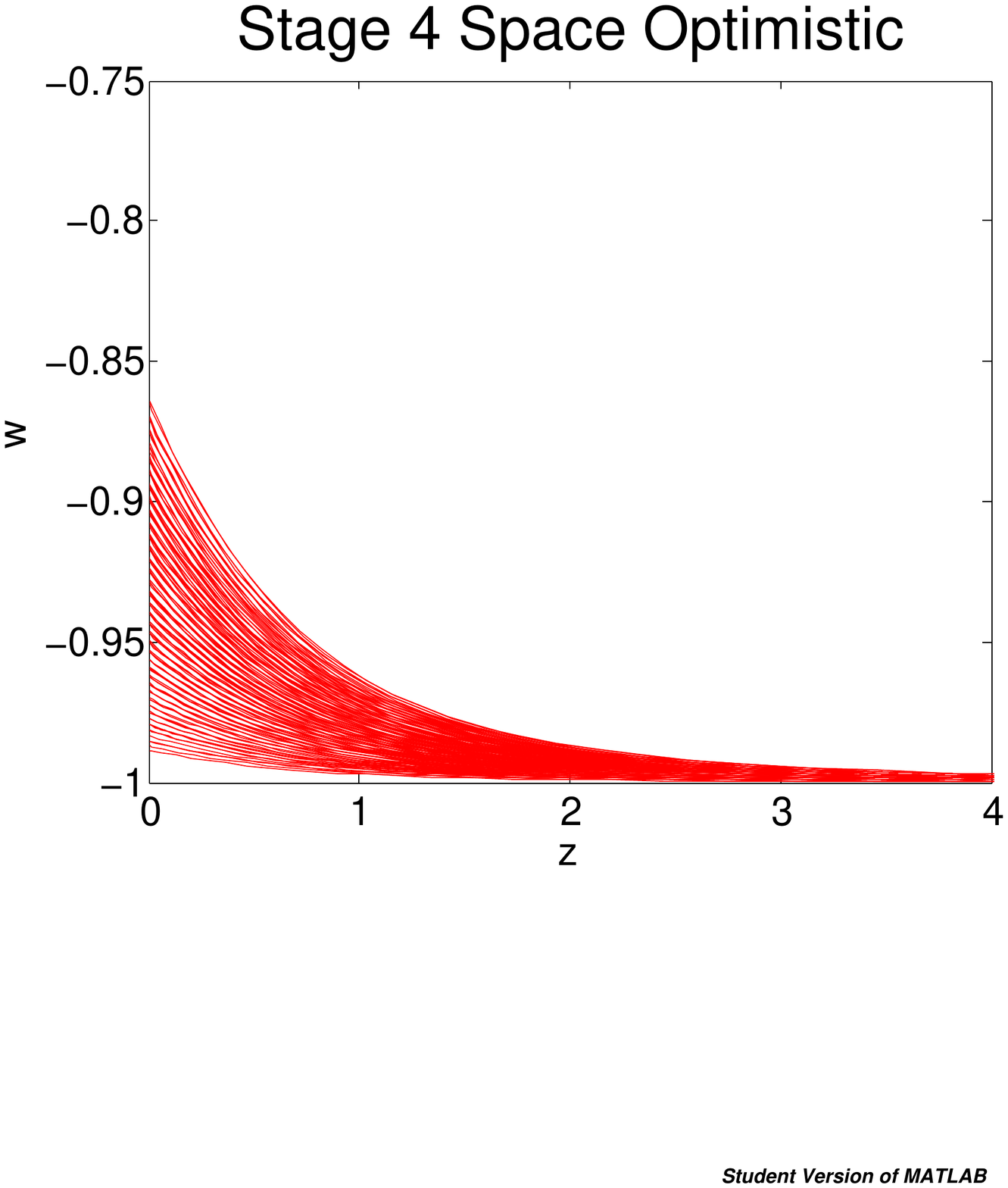}
\vspace{30pt}

\caption{The $w(z)$ behavior for a sample of points covering the full range of the $\lambda-V_I$ space for data based on an exponential model.}
\label{fig:ExpW(z)}

\end{figure*}

With our parameterization firmly in hand, we now analyze the DETF data
sets based on a cosmological constant cosmology using the MCMC technique. The
likelihood contours for Stage 2, Stage 3 Photo Optimistic, Stage 4 Ground LST
Optimistic, and Stage 4 Space Optimistic are shown in Fig. \ref{fig:lambdafid-pars} for the
$\lambda-V_I$ space.  In all plots shown in this paper we use the DETF
supernova, weak lensing, baryon oscillation, and PLANCK (using the alternate parameters as in \cite{Albrecht:2007qy}) data sets  but not the cluster data sets due to technical problems adapting the DETF
cluster data models to our methods. 
%We also use the alternate PLANCK parameter space used in \cite{Albrecht:103003}. 
These technical problems are similar to those outlined in \cite{Albrecht:2007qy}, where similar
issues where encountered.  Our plots were constructed by marginalizing
over all the cosmological parameters, $\omega_m$, $\omega_k$, $\omega_B$,
$\delta_\zeta$, $n_s$, the various nuisance parameters, and/or the photo-z
parameters. The nuisance and photo-z parameters are detailed in the
appendix of one of our companion papers\cite{Abrahamse:2007ah}. The fiducial values for
the cosmological parameters are shown in Table \ref{tab:Table1}.  The
values for all energy densities and $V_I$ in the remainder of the paper are in units of $h^2$, where
$h = \frac{H}{100}$. 

Figure \ref{fig:lambdafid-delta} gives likelihood contours in
$\lambda - \delta{\omega_{DE}}$ space, where $\delta{\omega_{DE}}
\equiv \omega_{DE}(a=1) - \omega_{DE}(a_I)$. Here $\omega_{DE} =
\frac{\rho_{\phi}}{\rho_{c}}h^2$ and $\rho_c = 3M_{p}^2H^2$. The value
of $\delta \omega_{DE}$ gives the amount the dark energy density
has changed since the simulation started at scale factor $a_I$ (well
in the radiation era). Values of $\delta \omega_{DE}$ different from zero
correspond to dynamical dark energy. Figure \ref{fig:lamfid-delta-enlarge} shows an enlarged version of the $\lambda - \delta{\omega_{DE}}$ space for Stage 3 and Stage 4 experiments.

%The plots in $\lambda-\delta{\omega_{DE}}$ space show the values of 
%lambda that correspond to an evolving dark energy. 
For Stage 2, values of $\lambda$ from 0 to about 0.15 in Fig. \ref{fig:lambdafid-delta} correspond to shallow slopes and
don't allow for much change in the amount of dark energy. For
these values of $\lambda$ there is a  spread in $V_I$ in the
$\lambda-V_I$ space (Fig. \ref{fig:lambdafid-pars}).  Since these values are consistent with a
non-evolving dark 
energy, the spread in $V_I$ is essentially a measure of how well the
experiment is measuring $\omega_{DE}(a=1)$. Values of $\lambda > 0.15$
all correspond to detectable differences from a cosmological
constant. This portion of the $\lambda-V_I$ space has an upturned
feature. As the slope gets steeper the field needs to start higher up
in the potential in order to roll down to acceptable values of
$\omega_{DE}(a = 1)$. These features are evident in the plots for
Stage 3 and Stage 4, although by Stage 4 it is less clear as the
parameter space has shrunken to values closer to a cosmological
constant, and so the upturned trend has diminished. 

Comparing Stage 2 to Stage 3 Photo Optimistic, and then on to Stage 4, there is a significant
tightening of the allowed area in parameter space. This increased
constraining power is similar to the factors of about three (Stage 2 to Stage 3) and ten
(Stage 2 to Stage 4) increase in constraining power
noted by the DETF in the $w_0-w_a$ space.
The $\lambda-\delta{\omega_{DE}}$ contours in
Fig. \ref{fig:lambdafid-delta} allow one to interpret the constraining
power already seen in  the  $\lambda-V_I$ space in terms of a specific
aspect of the  dark energy dynamics, namely the overall change in dark energy
density (given by $\delta \omega_{DE}$).

Figure \ref{fig:LambdaW(z)} shows plots of the allowed functions of
$w(z)$ for each data set. The plots are constructed by
selecting around 100 points taken uniformly within the $3\sigma$
contour (thus also including points within the $1$ and $2\sigma$
contours). This is done to illustrate the full range of solutions not
excluded at better than $3\sigma$. 
The furthest most curve from $w = -1$ for each Stage corresponds to
the top right most tip in the $\lambda-V_I$ space.

Since we are interpreting the data using the exponential quintessence
model, we can use our knowledge of how the cosmological solutions vary
with $\lambda$ to discuss constraints on future cosmology as
well. Since all the stages depicted in Figure
\ref{fig:lambdafid-pars} strongly favor $\lambda < \sqrt{2}$,
essentially all future $w(a)$ behavior consistent with these data sets will approach scaling
solutions where $w \rightarrow \frac{\lambda^2}{3} - 1 < -\frac{1}{3}$ and give
accelerating universes where $\Omega_{\phi} \rightarrow 1$. The
scaling solutions have not been reached by today but will be
approached in the future. Therefore, by Stage 2, solutions that lead
to universes with a non-accelerating fate have been ruled out in this scenario. We will
revisit this point in the next section in the context of a different
background cosmology.

\section{\label{sec:Sec5}Exponential Model Fiducial Data}

Next we consider the case where the universe happens to have dark energy
described by the exponential model.  We select a particular fiducial
model of dark energy to illustrate the potential impact of Stage 4
data. We use exponential model parameters of $\lambda = 0.7$
and $V_I = 0.42$ for the fiducial model (the remaining parameters were
given the same values we use for the cosmological constant background model, given in Table \ref{tab:Table1}).

We choose the fiducial values by finding a point in the
$\lambda-V_I$ space in Fig. \ref{fig:lambdafid-pars} for
Stage 2 that was just outside a 1$\sigma$ detection
but was excluded by better than 3$\sigma$ in Stage 4 Optimistic (for both ground and space). This
point corresponds to
$w(a=1) = -0.92$, as shown in Fig. \ref{fig:firstpot}.
The results of Stage 2, Stage 3
Photo Optimistic, Stage 4 Space Optimistic and Stage 4 LST Optimistic are shown in Figure
\ref{fig:Expfid-pars} for the $\lambda-V_I$ space. The $\lambda -
\delta{\omega_{DE}}$ space is shown in Fig. \ref{fig:Expfid-delta}. Fig. \ref{fig:Expfid-delta-enlarge} shows an enlarged version of the $\lambda - \delta{\omega_{DE}}$ space for Stage 4 experiments.

For Stage 2, the likelihood contours in $\lambda-V_I$ space look
similar to the case with a cosmological constant background
cosmology. 
The range of $\lambda$ is nearly the same with all $\lambda <
\sqrt{2}$. Therefore the same conclusions about strongly favoring
future scaling behavior that were made
in the previous section apply here. The upturned trend is a little
more dramatic since dark energy solutions with more evolution are
favored. The major difference in the space is that values of $\lambda
\lesssim 0.05$ are outside the 1$\sigma$ contour. Comparing this with the
$1\sigma$ region of $\delta{\omega_{DE}}$ shows that these values are
consistent with a non-dynamical dark energy. It is not until we reach
values of $\lambda \approx 0.2$ that we find a 1$\sigma$ region that
corresponds to an evolving dark energy in the $\delta{\omega_{DE}}$
plot. Therefore, even though the parameters consistent with a
cosmological constant fall outside the 1$\sigma$ contour in the
$\lambda-V_I$ plot, a non-dynamical dark energy, and thus a cosmological constant, is not
ruled out at even the 1$\sigma$ level by analyzing the contours in the
$\lambda-\delta w_{DE}$ space. This apparent discrepancy between the two
pictures is in fact a common situation when examining relatively low
likelihood contours in different parameter spaces. More clear signals
will only be found when looking at phenomena that are rejected at
higher likelihood levels.

\begin{table}[ht]
\centering
\caption{$\Lambda$CDM (left column) and Exponential (right column)
  Fiducial Parameter Values. The nuisance and photo-z parameters are
  0.} 
\begin{tabular}{|l l | c | r l|}
	\hline \hline
$\omega_{DE}$ & $0.3796$ & $0.3796$ \\
$\omega_{m}$ & $0.146$ & $0.146$   \\
$\omega_{k}$ & $0.0$ & $0.0$     \\
$\omega_{B}$ & $0.024$ & $0.024$  \\
$n_s$        & $1.0 $   & $1.0 $   \\
%$n_s'$       & $0.00001$ & $0.00001$ \\
$\delta_{\zeta}$  & $0.87$ & $0.87$  \\
$\lambda$  & $0.0$ & $0.7$  \\
$V_I$  & $0.38$ & $0.42$  \\
	\hline
\end{tabular}
\label{tab:Table1}
\end{table}

By Stage 3 Photo Optimistic, the cosmological constant is now excluded
in the 1$\sigma$ contour although well within the 2$\sigma$
contour in the $\lambda-V_I$ space. This is consistent with the $\lambda -
\delta{\omega_{DE}}$ plot in Fig. \ref{fig:Expfid-delta}. The
increased constraining power is again equivalent to the DETF result
for the $\lambda-V_I$ space. However, the 
range of $\lambda$ has not changed much within the 3$\sigma$ contour,
allowing the range of evolving dark energy solutions to be nearly as
large as Stage 2, as seen in Fig. \ref{fig:Expfid-delta}. 

Stage 4 clearly differentiates between the exponential fiducial model
and a cosmological constant by better than 3$\sigma$, as shown in both
$\lambda - \delta{\omega_{DE}}$ and $\lambda - V_I$ spaces depicted
in Figures \ref{fig:Expfid-pars} and \ref{fig:Expfid-delta}. Again,
the increased constraining power is consistent with the DETF results
for both Stage 4 experiments. 

Figure \ref{fig:ExpW(z)} shows the span of the $w(z)$ 
solutions for each of the Stages determined in the same way as in the
previous section. The curve with the greatest $w(z = 0)$ departure
from $-1$ is obtained from the top right tip of the $3\sigma$ curve in
the $\lambda-V_I$ space. For the Stage 4 plots, the bottom most curve,
corresponding to the closest approach to a cosmological constant,
corresponds to the bottom left tip of the $3\sigma$ region in the
$\lambda-V_I$ space.

\section{Discussion and Conclusions}

We have analyzed the exponential scalar field model using MCMC
techniques for the DETF simulated data sets representing future dark energy 
experiments. We have demonstrated the ability of these experiments to
place significant constraints on the parameters of a scalar field
model. The relative constraints on the size of the $\lambda-V_I$ space
between various data sets produce values similar to the constraints
computed by the DETF in the $w_0-w_a$ space. 
In addition to placing constraints on the quintessence parameters, we 
also presented our results in terms of the evolution of the dark
energy in our quintessence model. This allows us to distinguish more
directly between a cosmological constant and our particular scalar field
model. We have
presented plots for a characteristic selection of combined DETF data
models, but in the course of this work we have also examined similar
plots for a much wider range of DETF data models, including data models
representing single techniques.  We found that the consistency with
constraints reported by the DETF in $w_0-w_a$ space to hold across the
entire range of data choices and combinations we considered.

We based our data on two different background cosmologies, one with a
cosmological constant and one with exponential model quintessence with
specific parameters. We found that the 
equivalence with the DETF results held in both cases.  
Our specific background quintessence model was chosen (with the parameter values of $V(\phi_I) = 0.42$ and
$\lambda = 0.7$) in order to illustrate the power of Stage 4
experiments.  For this model, the maximum deviation from $w=-1$ occurs
today, with  $ w(a=1) = -.92$. 
We found that if the universe is accelerating due to this particular
exponential quintessence model 
then a cosmological constant dark energy model can be ruled out to at
least 3$\sigma$ by good Stage 4 experiments. For this background
cosmology, the cosmological constant is within the 1$\sigma$
contour at Stage 2 and the 2$\sigma$ contour at Stage 3. 

We note that there are a number of ways experiments might be optimized
to do better than the cases considered by the DETF (see for example
\cite{Zhan:2006gi, Schneider:2006br}). We have not included such
ideas in our work, with an eye for offering more direct points of
comparison with the DETF.  However, improvements such as these could
lead to more powerful constraints on quintessence models than we have
calculated here. 

We have found in this work and in our companion papers
\cite{Abrahamse:2007ah,Barnard:2007ah} that a wide variety of quintessence models (with
widely varying families of functions $w(z)$) are
constrained by DETF data in a way comparable to the constraints
found in $w_0-w_a$ space by the DETF.  As discussed in \cite{Albrecht:2007xq}, we believe that this is related to
recent work by one of us (AA) and Bernstein \cite{Albrecht:2007qy},
where it was demonstrated that, overall, the good DETF simulated data sets could
constrain significantly more than two dark energy parameters.  From
this point of view our various models of dark energy are just
sampling different more or less ``random'' combinations of the ``well measured modes''
discussed in \cite{Albrecht:2007qy} and in each case are coming up with similar
results.

\begin{acknowledgements} We thank
  Matt Auger, Lloyd Knox and Michael Schneider for
  useful discussions and constructive criticism. Thanks also to Jason
  Dick who provided much useful advice on MCMC. We thank the Tony
  Tyson group for use of their computer cluster, and in particular
  Perry Gee and Hu Zhan for expert advice and computing support. Gary
  Bernstein provided us with Fisher matrices suitable for
  adapting the DETF weak lensing data models to our methods, and David
  Ring and Mark Yashar provided additional technical assistance.  This work was
  supported by DOE grant DE-FG03-91ER40674 and NSF grant AST-0632901. 
\end{acknowledgements}

\bibliography{ExpBibFinal}

\end{document}